\documentclass[sigconf]{acmart}

\usepackage[whole]{bxcjkjatype}

\usepackage{fancybox}
\usepackage{lscape}
\usepackage{makecell}
\usepackage{multirow}
\usepackage{subcaption}
\usepackage{float}
\usepackage{soul,color}
\usepackage{colortbl}

\AtBeginDocument{%
  \providecommand\BibTeX{{%
    \normalfont B\kern-0.5em{\scshape i\kern-0.25em b}\kern-0.8em\TeX}}}

\setcopyright{rightsretained}
\copyrightyear{2024}
\acmYear{2024}
\acmDOI{10.1145/3640794.3665549}

\acmConference[CUI '24]{ACM Conversational User Interfaces 2024}{July 8--10, 2024}{Luxembourg, Luxembourg}
\acmBooktitle{ACM Conversational User Interfaces 2024 (CUI '24), July 8--10, 2024, Luxembourg, Luxembourg}
\acmISBN{979-8-4007-0511-3/24/07}

\begin{document}

\title[Coaching Copilot: Blended Form of an LLM-Powered Chatbot and a Human Coach ...]{Coaching Copilot: Blended Form of an LLM-Powered Chatbot and a Human Coach to Effectively Support Self-Reflection for Leadership Growth}

\author{Riku Arakawa}
\orcid{0000-0001-7868-4754}
\affiliation{%
  \institution{Carnegie Mellon University}
  \city{Pittsburgh}
  \country{USA}
}
\email{rarakawa@cs.cmu.edu}
\authornote{These authors contributed equally and are ordered alphabetically.}

\author{Hiromu Yakura}
\orcid{0000-0002-2558-735X}
\affiliation{
    \institution{Max-Planck Institute for Human Development}
    \city{Berlin}
    \country{Germany}
}
\email{yakura@mpib-berlin.mpg.de}
\authornotemark[1]

\renewcommand{\shortauthors}{Arakawa and Yakura}


\newcommand{\tabref}[1]{Table~\ref{#1}}
\newcommand{\figref}[1]{Figure~\ref{#1}}
\newcommand{\secref}[1]{Section~\ref{#1}}
\newcommand{\eqnref}[1]{Equation~\ref{#1}}
\newcommand{\appref}[1]{Appendix~\ref{#1}}

\newcommand{\eg}{\textit{e.g.},~}
\newcommand{\ie}{\textit{i.e.},~}
\newcommand{\etc}{\textit{etc}.}
\newcommand{\etal}{\textit{et al.}}

\newcolumntype{C}[1]{>{\centering\arraybackslash}m{#1}}


\begin{abstract} 
Chatbots' role in fostering self-reflection is now widely recognized, especially in inducing users' behavior change.
While the benefits of 24/7 availability, scalability, and consistent responses have been demonstrated in contexts such as healthcare and tutoring to help one form a new habit, their utilization in coaching necessitating deeper introspective dialogue to induce leadership growth remains unexplored.
This paper explores the potential of such a chatbot powered by recent Large Language Models (LLMs) in collaboration with professional coaches in the field of executive coaching.
Through a design workshop with them and two weeks of user study involving ten coach-client pairs, we explored the feasibility and nuances of integrating chatbots to complement human coaches. 
Our findings highlight the benefits of chatbots' ubiquity and reasoning capabilities enabled by LLMs while identifying their limitations and design necessities for effective collaboration between human coaches and chatbots.
By doing so, this work contributes to the foundation for augmenting one's self-reflective process with prevalent conversational agents through the human-in-the-loop approach.
\end{abstract}

\ccsdesc[500]{Human-centered computing~Interaction design}
\ccsdesc[500]{Human-centered computing~Collaborative and social computing}

\keywords{coaching, reflection, chatbot, human-AI collaboration}

\maketitle

\section{Introduction}
\label{sec:intro}

Self-reflection is an indispensable element for personal growth and behavior change, and research to foster it is actively conducted in the Human-Computer Interaction (HCI) field~\cite{DBLP:conf/ACMdis/BaumerKMRSG14, DBLP:journals/imwut/AlqahtaniJV20, DBLP:journals/imwut/BentvelzenWHSN22}.
In this field, chatbot-based coaching has been attracting attention with a potential to enable 24/7 support for reflection, the effectiveness of which has been demonstrated in areas like healthcare~\cite{DBLP:journals/imwut/KocielnikXAH18, DBLP:journals/pacmhci/MitchellMCTDSM21}, well-being support~\cite{DBLP:conf/huc/KamaliACAKM18}, and education~\cite{DBLP:conf/hci/MaiWRP21}.
While common in forming a new habit (\eg healthy diet, exercise), using chatbots for supporting the achievement of professional goals, such as leadership growth in an organization, is less covered in prior research.
Notably, such goal achievement demands a nuanced approach to fostering deep, introspective analysis of one's own behaviors and performing strategic decision-making.
This paper explores the effective use of chatbots to facilitate such deep introspective reflection to augment the potential of today's conversational agents, especially in the paradigm of Large Language Models (LLMs).

In practice, \textit{executive coaching} has played a significant role in talent development for many years~\cite{witherspoon1996executive, doi:10.1177/0149206305279599}.
Executive coaching typically consists of a dialogue between a coach and their client, where the coach attempts to bring out the leadership qualities of a client through dialogues, and often involves goal setting, action planning, and accountability for professionals in high-stakes, high-stress roles~\cite{Kilburg1997}.
The unique challenges lie in navigating complex interpersonal dynamics, understanding the subtleties of organizational cultures, and facilitating transformative insights that can guide high-level strategic decisions while maintaining clients' motivation to change behaviors~\cite{Joo2005Executive}. 
As such, in recent years, there has been increasing attention in both HCI~\cite{DBLP:conf/chi/ArakawaY19, DBLP:conf/chi/ArakawaY20} and human resource development (HRD)~\cite{grassmann2021coaching, weimann2022virtual, Bridgeman2023Using} on research to realize technological support for executive coaching.

We took executive coaching as a field to explore how LLM-powered chatbots can effectively foster one's reflection to achieve professional goals.
This was motivated by prior work in HRD~\cite{grassmann2021coaching}, which pointed out that merely introducing a chatbot in executive coaching encounters difficulties in identifying clients' core problems and providing individual, precise feedback.
The work suggests the need to design the use of chatbots carefully with domain experts. 
Therefore, in this paper, we conducted two studies in collaboration with coaching organizations.
First, we held a workshop with eight professional coaches, where they were asked to interact with a GPT-4-based chatbot prototype while imagining using it for coaching processes.
Through the discussion, we found that the technology would be best suited to offering complementary text coaching between face-to-face sessions.
We then developed an LLM-powered chatbot dedicated to such a blended experience. 
We conducted a two-week user study with ten pairs of coaches and clients in actual coaching situations.
The semi-structured interviews with both coaches and clients after the trial shed light on the advantages and limitations of chatbot coaches, as well as the importance of securing commitment with human coaches.
The study supplementarily showed that the clients' behavioral intention was sustained, and their authenticity scale was affected throughout the trial.
These results confirmed the benefit of conversational agents in complementing existing coaching practices for leadership growth.

Based on the study's implications, we summarized a guideline to support one's achievement of their professional goals by blending an LLM-powered chatbot coach and a human coach.
While prior work suggested the necessity of the collaboration between human and artificial intelligence (AI) in chatbot-based reflection support~\cite{grassmann2021coaching}, actual collaboration practice has not been well-established, particularly in such a blended manner.
Our work offers a set of key factors with empirical evidence guided by both coaches and their clients in actual executive coaching sessions.
Furthermore, these findings distinctly delineate the current performance boundaries of LLM-powered chatbots in fostering deep self-reflection, which demands a high level of human context understanding, and hint at the future directions of human-centered natural language processing and conversation analysis research.

\section{Background}
\label{sec:bg}

Our research intersects with various aspects of both HRD practice and HCI research.
To begin with, we offer foundational background information to set the context.
We then review prior work on introducing a conversational agent to support different kinds of reflection to situate our work.

\subsection{Executive Coaching and Technological Support}
\label{sec:bg-executive}

Executive coaching is a professional development process in HRD designed to enhance the performance of individuals in executive or leadership positions within organizations~\cite{witherspoon1996executive, doi:10.1177/0149206305279599}.
This is achieved through a form of dialogue where a coach works with executives or managers in a business to help them improve their leadership skills, gain self-awareness, clarify their goals, and achieve their development objectives~\cite{Kilburg1997}.
To this aim, typical coaching heavily relies on deep, spontaneous reflection regarding one's behaviors, beliefs, and values, leading to a clearer understanding of one's own motivations, strengths, and areas for improvement~\cite{moen2009effect}.
A characteristic point of executive coaching is that the dialogue is not only oriented toward immediate problem-solving or providing feedback but focused on guiding the client's self-reflection through questions~\cite{correia2016understanding}.
Therefore, the coach's skills are essential for successful outcomes by fostering clients' readiness~\cite{Grant2012Making}, facilitating their reflection, and maintaining their motivation~\cite{MacKie2015Effects}.
While getting soaring attention, the industry faces the problem of lacking proficient coaches, preventing the practice from being affordable for everyone who needs it~\cite{Hawkins2008}.

To make such coaching processes scalable, several technological supports have been proposed in HCI research, exploring the potential of computers to assist the process from the perspective of human-AI interaction~\cite{DBLP:conf/chi/AmershiWVFNCSIB19}.
For example, \textit{REsCUE}~\cite{DBLP:conf/chi/ArakawaY19} is a real-time feedback system for a coach during coaching sessions based on clients' multimodal behavior signals such as gaze.
It was suggested that such feedback could deepen the ongoing conversation quality from the coach's perspective.
\textit{INWARD}~\cite{DBLP:conf/chi/ArakawaY20} is a video-reflection supporting system for both coach and client, successfully making the reflection process more efficient and effective for them.
These systems were designed to support difficult or tedious processes within executive coaching (\eg watching an entire video of the coaching session) while preserving the form of the existing coach-client communication.
To further broaden coaching reach and cater to a larger audience in need, this study explores how intelligent conversational agents powered by recent advancements in LLMs can augment the coach-client communication.

\subsection{Reflection Support in HCI}
\label{sec:bg-reflection}

Meanwhile, reflection is a long-studied, multi-faceted concept given its importance in personal growth and behavior change~\cite{DBLP:conf/ACMdis/BaumerKMRSG14, DBLP:journals/imwut/BentvelzenWHSN22}.
HCI research has tried to support users' reflection with different types of artifacts (\eg virtual reality application~\cite{DBLP:conf/chi/WagenerRZBSWBSM23}, avatar coach~\cite{DBLP:journals/computer/ZhaoLQL20}, \etc).
Examples span a wide range of contexts such as personal informatics~\cite{DBLP:conf/iui/JorkeSMSR23}, health and well-being support~\cite{DBLP:journals/pacmhci/MitchellMCTDSM21, DBLP:conf/huc/KamaliACAKM18, DBLP:conf/chi/RyanDL22}, parent-child interaction~\cite{DBLP:journals/imwut/KimLKJYHKS20}, school education~\cite{DBLP:conf/hci/MaiWRP21}, professional learning~\cite{DBLP:journals/tlt/WolfbauerPMR22}, video meeting~\cite{DBLP:conf/cscw/BenkeVM21,DBLP:conf/chi/SamroseMSSRHRMC21,DBLP:journals/corr/abs-2402-11145}, and creativity support~\cite{DBLP:conf/chi/0002B23}.
For example, Mai~\etal~\cite{DBLP:conf/hci/MaiWRP21} examined the efficacy of chatbot in helping students reflect on their exam anxiety and found the benefit of computer-driven opportunity in reducing the hurdle for students to initiate reflection.
Wolfbauer~\etal~\cite{DBLP:journals/tlt/WolfbauerPMR22} presented that a guidance system to support reflective writing can enhance a user's reflection competence.
These various reflection-support systems were recently reviewed by Bentvelzen~\etal~\cite{DBLP:journals/imwut/BentvelzenWHSN22}, who concluded that the level of reflection is diverse in different contexts, calling for more empirical evidence with a situated evaluation of the reflection effect.
Our research aims to contribute to the ongoing discussions by providing an example where we investigate the design and effectiveness of reflection-support technology with a strong focus on leadership growth, which requires unique, deep introspection.
Considering its dialogic essence to foster reflection in contrast to methods like reflective writing via fixed questions~\cite{Moussa-Inaty2015}, executive coaching would be particularly well-suited to explore the capabilities of LLM-powered chatbots.

\subsection{Chatbot-Based Coaching}
\label{sec:bg-chatbot}

Dialogue can play a key role in our reflection process~\cite{DBLP:journals/imwut/BentvelzenWHSN22}, as Mols~\etal~\cite{Mols2016Informing} found that most people conduct reflection while conversing with others.
Thus, interactive chatbots have been studied as a medium to facilitate one's reflection~\cite{Flstad2019Different}, aiming to be a virtual coach in multiple areas.
Specifically, the possibility has been widely examined in healthcare~\cite{DBLP:conf/chi/MitchellEM22, DBLP:journals/pacmhci/MitchellMCTDSM21, DBLP:conf/chi/RutjesWI19, DBLP:conf/chi/QiuYBHY23, DBLP:conf/chi/RyanDL22, Beaudry2019Getting, Wlasak2023Supporting, Casas2018Food}, well-being support~\cite{Narain2020Promoting, Gabrielli2020, Xygkou2023Conversation, Cai2023Listen}, education~\cite{Terblanche2022Performance, Essel2022Impact, DBLP:conf/hci/MaiWRP21, DBLP:journals/eait/KuhailAAA23}, and team collaboration~\cite{DBLP:conf/cscw/BenkeVM21, DBLP:conf/chi/SamroseMSSRHRMC21}.
For example, Mitchell~\etal~\cite{DBLP:journals/pacmhci/MitchellMCTDSM21} provided empirical evidence that a script-based chatbot can support individuals with type 2 diabetes.
They concluded that the benefit of the chatbot coach is its persistence and consistency, fostering clients' autonomy.
Similarly, Essel~\etal~\cite{Essel2022Impact} reported the chatbot's effectiveness in undergraduate students' learning behavior and confirmed its benefit in responding swiftly to the questions the students asked.
Gabrielli~\etal~\cite{Gabrielli2020} presented a similar result in helping adolescents gain coping skills and mental well-being.
These studies demonstrate the power of the chatbot that is always available and not affected by external factors like humans (\eg by fatigue or unrelated emotional events) in its responses.

As discussed in \secref{sec:bg-executive}, executive coaching has a unique practice of having reflection through dialogue, and these benefits of chatbot may not be directly applicable.
For instance, recent work in HRD argued that replacing human coaches with AIs is not optimal due to its insufficient capability to deal with diverse dialogue contexts, and the technology may work better for a set of specified topics in executive coaching~\cite{grassmann2021coaching}.
Terblanche and Cilliers~\cite{Terblanche2020Factors} concluded that what ``works'' in human-to-human coaching may not necessarily be applicable in AI-to-human coaching because users of a chatbot coach may struggle to trust an algorithm.
Passmore and Tee~\cite{Passmore2023} mentioned that, despite chatbots' scalability, their inability to demonstrate empathy and emotional intelligence could be critical limitations in this sensitive domain.
These papers call for empirical substantiation of a better approach to incorporating AIs than trying to alternate human coaches with chatbots.

Our work is inspired by a position paper of Gra{\ss}mann and Schermuly~\cite{grassmann2021coaching} that implied the potential of augmenting traditional coaching with conversational agents rather than replacing human coaches, given the above limitations.
Specifically, in this work, we seek a plausible form of blending a human coach and a chatbot coach, especially in the paradigm of the rapidly developed LLM-powered chatbots~\cite{Dwivedi2023Opinion}.
To this end, we distinctly define the capacities and roles of each party and scrutinize their impact through collaborative design with domain experts.
We note that such an approach of blending a human and a chatbot coach is not yet prominent.
Thus, we believe that our design considerations rooted in this unique domain can also have implications for the HCI community to diversify the ways of leveraging chatbots.

\section{Research Questions}
\label{sec:rq}

This paper explored the effective use of LLM-powered chatbots to induce deep self-reflection and behavior change beyond habit formation.
For this, we took executive coaching as a study field and focused on one's leadership growth.
Throughout the rest of this paper, we seek answers to the following research questions.
First, we want to discern the potential areas within the stages of the executive coaching process where chatbots can play a contributive role.
\begin{quote}
    \textbf{RQ1}: What part of executive coaching can LLM-powered chatbot get involved with?
\end{quote}
\begin{quote}
    \textbf{RQ2}: What do human coaches consider as the expected role of such a chatbot and their own for successful coaching?
\end{quote}

Then, as highlighted in \secref{sec:bg-reflection}, there is a need for empirical evidence to understand how technology aids in facilitating individual reflection.
Hence, we will address the next question:
\begin{quote}
    \textbf{RQ3}: How do clients perceive the introduction of such an LLM-powered chatbot coach in their coaching processes and exploit it?
\end{quote}
By exploring these questions, we expected that we could derive insights into supporting professional coaches with LLM-powered chatbots in connection to prior research in HCI.

\section{Workshop with Professional Coaches}
\label{sec:workshop}

To answer \textbf{RQ1} and \textbf{RQ2}, we conducted a design workshop with professional coaches in June 2023.
The purpose was to get their perspectives on the state-of-the-art LLM-powered chatbot concerning its capability and potential role in executive coaching.

\subsection{Procedure}

Eight coaches (two male and six female, 40--56 years old) were invited to our workshop via a Japanese company that offers executive coaching services to its client companies.\footnote{Here, we started the workshop by focusing on the coach side because we had difficulties inviting clients due to their contracts.}
An experimenter first conducted a pre-use hearing where they informally asked their thoughts about the potential of recent chatbots, such as ChatGPT, especially in leadership growth.
Then, the experimenter introduced the playground of GPT-4\footnote{\url{https://gpt4demo.com/apps/gpt4-playground}} and explained basic concepts of using LLM-powered chatbots, such as system prompting.
Here, the coaches freely edited the prompt and had a conversation with GPT-4.
After familiarizing themselves with the experience, the coaches discussed the possibility and difficulty of introducing such chatbots into executive coaching.
Also, the experimenter conducted semi-structured interviews with them, asking their opinions about the discussion topics and their prospect about the role and value of human coaches after introducing the chatbots.
All the above processes were conducted remotely with the attendance of the experimenter and recorded under their agreement.

\subsection{Findings}
\label{sec:workshop-findings}

In the pre-use hearing, the coaches mentioned various use cases potentially made possible by chatbots.
They can be mainly categorized into supporting face-to-face and text coaching sessions.
Regarding the former direction, the coaches mentioned the possibility of real-time assistance:
\begin{quote}
    I appreciate it if chatbots could provide us with a selection of potential questions based on transcribing and comprehending the content of the dialogue with clients. Even if the suggestions are not optimal, they could inspire us to formulate improved questions.
\end{quote}
On the other hand, most coaches expressed concerns about the capability of the chatbots, believing that it would be infeasible for them to replace face-to-face conversations fully:
\begin{quote}
    It requires many skills to deal with highly human context [sic], not only listening to their words but also monitoring their behavior. Just asking template questions would never facilitate their reflection.
\end{quote}
We understand this comment reflected the coach's limited trust in the chatbot's capability given the required skills in executive coaching as discussed in~\secref{sec:bg-executive}, although LLM-powered chatbots could go beyond simple template-based communication. 
Rather, they showed more positive attitudes to the chatbot assisting text communication between coaches and clients to reduce their burden and scale the coaching experience.
\begin{quote}
    I recognize the significance of text coaching in face-to-face sessions in maintaining relationships with clients and stimulating their actions. However, the difficulty lies in the fact that it could consume unlimited time, unlike face-to-face sessions with fixed schedules. AI coaches could adequately fulfill this role, possibly maintaining the quality and frequency of text communication.
\end{quote}
\begin{quote}
    Text communication between sessions is valuable to build a good foundation and to keep the momentum of their reflection, but in reality, it is not always possible for us to keep texting them, which could be counterproductive.
\end{quote}
From these comments in the pre-use hearing regarding \textbf{RQ1}, we found that human coaches should keep playing a major role in facilitating clients' reflection to deal with diverse and highly interpersonal contexts in executive coaching.
We also found that using chatbots as a text coach between face-to-face sessions to supplement them would be a promising direction to assist the coaches; although such communication is critical, it is a significant labor for the coaches, and clients could lose motivation by insufficient reply.
Such a collaborative direction was further supported by the fact that coaches emphasized human coaches' importance for clients' behavior change motivation.
\begin{quote}
    One of the reasons coaching is effective is the sense of being invested time by other people, so I believe fully automating with AI would be challenging.
\end{quote}
Similar comments were obtained in the semi-structured interviews after the coaches interacted with a GPT-4-based chatbot.
\begin{quote}
    The chatbot's response was better than I had expected. I believe it could be effective for people motivated to change their behavior independently, but for others, having a human coach to accompany them still seems crucial. For instance, if there is an opportunity to form a clear intention of how to and how often to use a chatbot coach beforehand through a dialogue with a human coach, then I think the client would achieve behavioral change.
\end{quote}
These comments suggested that preparing clients' readiness for the chatbot coach is critical where they gain the motivation to keep using the system and the perspective regarding how it can foster their reflection.
Moreover, the coaches agreed that human coaches should help the clients when they lose a way to utilize the chatbot coach.
\begin{quote}
    I can envision cases where clients might stop using it midway due to the lack of a sense that someone is watching over them. The kind of relationship where a human coach provides comments or offers support when they are stuck will likely be necessary.
\end{quote}
This comment emphasizes the importance of a human coach's presence for sustainable engagement, suggesting that it would not be optimal to fully automate the process even with common computational intervention approaches such as auto-reminders. 
The coaches concluded that their guidance on fostering the clients' readiness to use chatbot coaches for behavioral change and maintaining functional relationships with chatbot coaches would be critical.
This informed us of the answers to \textbf{RQ2}; blending coaching in collaboration of human and chatbot coaches can open up the way to providing successful coaching sessions in a scalable manner by reducing the workload of text coaching of human coaches.

\begin{figure*}[t]
    \centering
    \includegraphics[width=\textwidth]{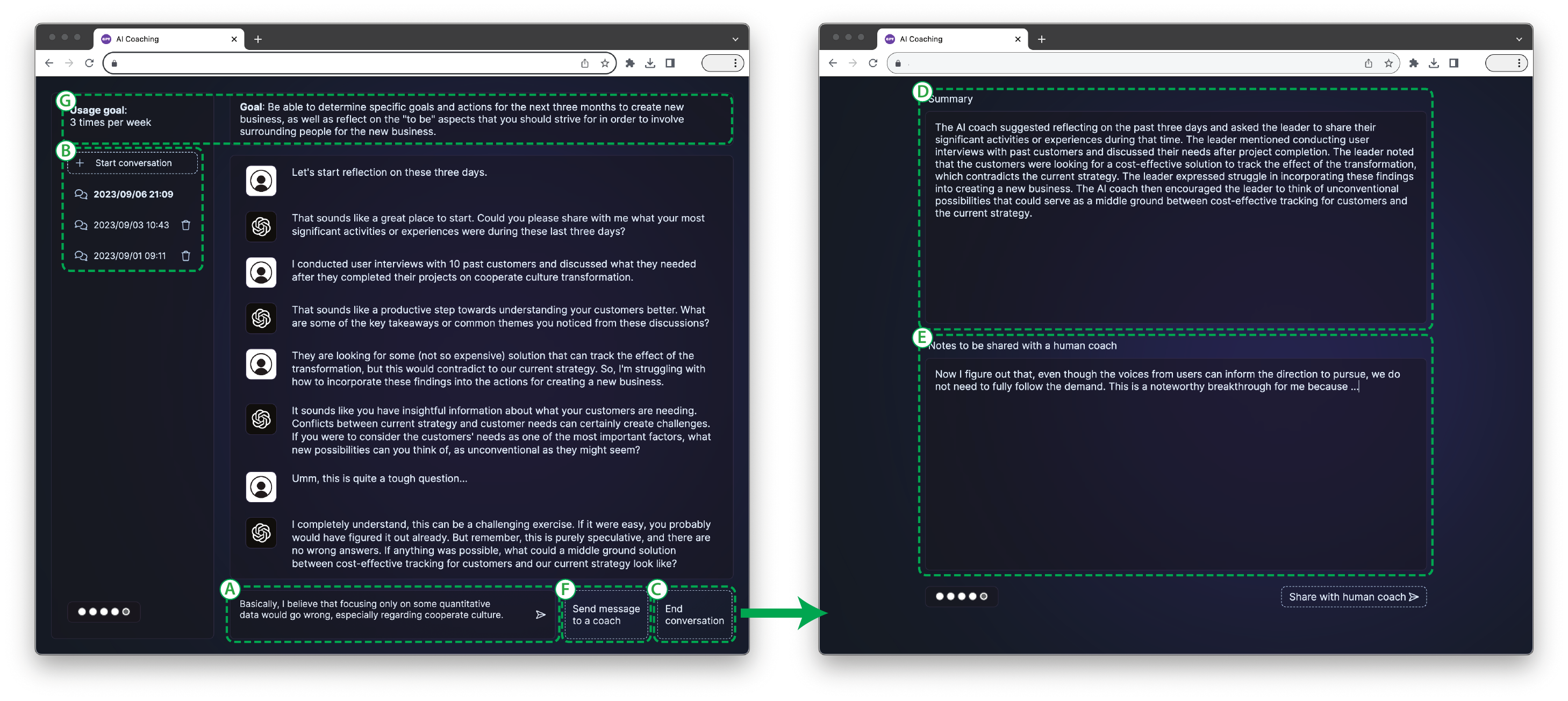}
    \caption{The prototype interface to use an LLM-powered chatbot coach. In the user study, this prototype was provided to clients and used at their own pace for two weeks to support their reflection to accomplish their professional goals. The presented conversation was derived from the use of a prototype by one of the authors to demonstrate its behavior.}
    \Description{The two screenshots of the prototype are presented. On the left one, A) a textarea to send a message, B) a list of past conversations, C) a button to end the conversation, F) a button to call a human coach, and G) a text of the declared goal are presented. In the right one, D) a textarea that contains the summary of the conversation and E) a textarea where a user can fill out comments are presented.}
    \label{fig:prototype}
\end{figure*}

\section{Evaluation Study: Design and Method}
\label{sec:method}

Based on the findings in the workshop, we prepared a chatbot-based prototype system for text coaching to empirically examine its influence on the clients in ecologically valid scenarios.
We deployed a system in actual coaching scenes and tracked clients' reflections and interactions with the system over two weeks during July and August 2023.
Finally, we conducted interviews with both coaches and clients to investigate the experience brought by the system qualitatively and to draw implications.

\subsection{Prototype}
\label{sec:method-prototype}

\figref{fig:prototype} presents the system we developed for the study.
Informed by the workshop, this prototype focused on text coaching and was implemented to enable clients to have a text chat with the GPT-4-based\footnotemark{} chatbot coach (\figref{fig:prototype}A).
The chat space is created for each text coaching session (\figref{fig:prototype}B).
Clients can end the session (\figref{fig:prototype}C) and then automatically send a report to the human coach via email so that the coach can be involved in the clients' reflection process, as the workshop suggested.
Here, the clients can revise the summary (\figref{fig:prototype}D) and messages (\figref{fig:prototype}E) to the human coach, where the summary is first generated by GPT-4.
Moreover, we allow the clients to send a message to their human coaches in the middle of the text coaching with the chatbot coach (\figref{fig:prototype}F), which was guided by the comment in the workshop that the human coach would be required to help the client when needed.
We also designed a prompt for the chatbot to suggest contacting the human coach if they struggle to maintain the conversation.
\footnotetext{We used \texttt{gpt-4-0613} with \texttt{temperature = 0}.}

Importantly, we designed the prototype so that there would be a face-to-face session where a human coach would explain the use of the prototype and set a goal with the client before starting the chatbot-based text coaching.
This is based on the important finding obtained in the workshop; that is, clients' readiness is critical to foster one's reflection using the system.
We used this face-to-face session to set up their expectation during the trial period.
More specifically, after explaining the capability of the chatbot coach, the human coach helped the client write their expected goal through the text coaching process and how many times they would use the system during the trial period.
Note that the goal and usage expectation were set individually by each pair of the coach and client and were kept visible on the main page throughout the trial (\figref{fig:prototype}G).

\begin{figure*}[t]
    \centering
    \fbox{
    \begin{minipage}{0.85\textwidth}
    You are a professional executive coach. Your role is to enhance the client's self-awareness and bring about behavioral change through precise questioning and feedback in your interactions with them.\\
    \\
    The client's desired goals through coaching: \texttt{\{\{ goal \}\}}\\
    What the client expects from you: \texttt{\{\{ expectation \}\}}\\
    \\
    Please be mindful not to present multiple questions in a single interaction to avoid confusing the client. Furthermore, if you find it challenging to bring about behavioral change in your conversation with the client, please present the following message to them:\\
    Speaking with a human coach might help clarify your thoughts further. What do you think about using the ``Send message to a coach'' button?
    \end{minipage}
    }
    \caption{The prompt used in the chatbot guided by the workshop with coaches.}
    \Description{The image of the prompt text. The text starts by ``You are a professional executive coach. Your role is to enhance the client's self-awareness and bring about behavioral change through precise questioning and feedback in your interactions with them.''}
    \label{fig:prompt}
\end{figure*}

The information was also used to create a prompt for the LLM inside the chatbot, as shown in \figref{fig:prompt}.
Here, \texttt{\{\{\,goal\,\}\}} and \texttt{\{\{\,expectation\,\}\}} refer to the client's goal and expectation about the text coaching process, which were set during the face-to-face session and registered on the system after the session by the coach.
This prompt was built on top of what we learned from the coaches' trials and errors in the workshop; for example, the sentences explaining the role were obtained from those the professional coaches had used to make the LLM imitate their conversations.
As described above, we also allowed the LLM to suggest a client communicate with a human coach if needed.
In addition, we observed that the LLM often produced multiple questions in a single response.
According to the professional coaches, it was pointed out to lead to clients' confusion.
Thus, we specified the LLM not to present multiple questions at once and finalized the prompt after several testing iterations.

\subsection{Participants}

We first recruited coaches from two coaching companies.
We explained the concept of our study and asked them to introduce the prototype into their actual coaching with their clients.
As a result, ten pairs of coaches and their clients, who regularly have face-to-face sessions on a monthly or bi-weekly basis, participated in our study.
We refer to clients as Cl1 -- Cl10 and paired coaches as Co1 -- Co10.
The detailed background about the clients is summarized in~\tabref{tbl:clients}.

\begin{table}[t]
  \caption{Backgrounds of the clients who participated in the two-week user study.}
  \label{tbl:clients}
  \begin{tabular}{rcccl}
    \toprule
         & Gender & Age & History with the current coach \\
    \midrule
    Cl1  & M      & 50s & 1 year                         \\
    Cl2  & F      & 40s & 8 months                       \\
    Cl3  & F      & 40s & none                           \\
    Cl4  & F      & 40s & 3 months                       \\
    Cl5  & F      & 30s & 2 years                        \\
    Cl6  & M      & 40s & 1.5 years                      \\
    Cl7  & F      & 30s & 2 months                       \\
    Cl8  & F      & 50s & 4 months                       \\
    Cl9  & F      & 40s & none                           \\
    Cl10 & F      & 20s & 2 months                       \\
    \bottomrule
  \end{tabular}
\end{table}

\subsection{Metrics}

\begin{figure*}[t]
    \centering
    \includegraphics[width=\linewidth]{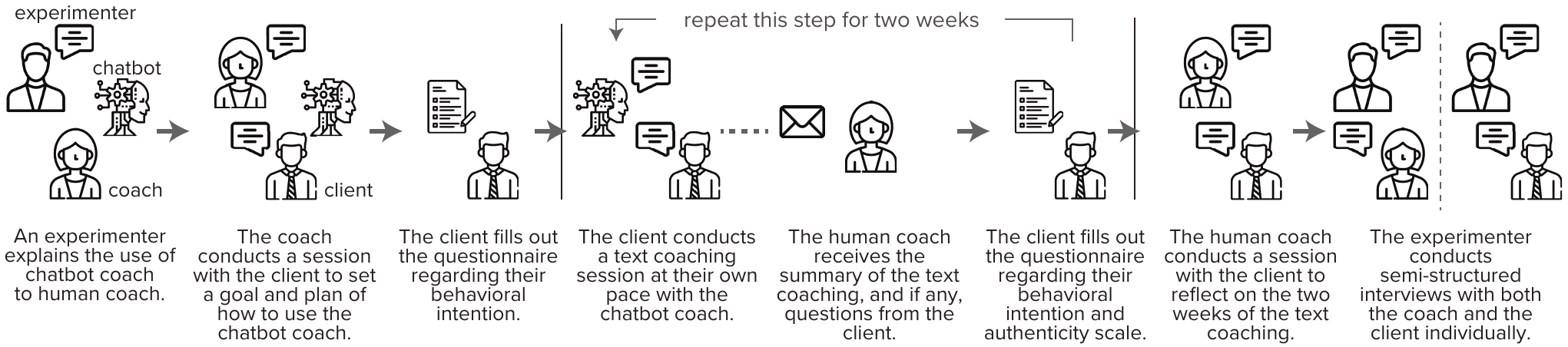}
    \caption{The procedure of the user study using the developed chatbot text coach as a supplement to the regular coaching.}
    \Description{The procedure of the study. 1) An experimenter explains the use of the chatbot coach to a human coach. 2) The coach conducts a session with the client to set a goal and plan for how to use the chatbot coach. 3) The client fills out the questionnaire regarding their behavioral intention. 4) The human coach conducts a session with the client to reflect on the two weeks of the text coaching. 5) The human coach receives the summary of the text coaching, and if any, questions from the client. 6) The client fills out the questionnaire regarding their behavioral intention and authenticity scale. 7) The human coach conducts a session with the client to reflect on the two weeks of the text coaching. 8) The experimenter conducts semi-structured interviews with both the coach and the client individually.}
    \label{fig:procedure}
\end{figure*}

To elucidate how the system affects the client, we used questionnaire-based measurements and asked the clients to fill them out every time they used the prototype.
In executive coaching, one's behavior or attitude to change shows a gradual transition, and we expected that investigating the transition using repeated questionnaires could provide insights.
Given their labor, we focused on two aspects, \ie the outcome of the reflection and attitude toward using the system, and used the corresponding two metrics.

\subsubsection{Authenticity Scale}

There are a couple of approaches for assessing the quality of one's reflection~\cite{DBLP:journals/imwut/BentvelzenWHSN22}.
The authenticity scale~\cite{Wood2008Authentic} has been used in the context of executive coaching~\cite{susing2011potential, DBLP:conf/chi/ArakawaY20} since it is about aligning one's inner feelings and primary experiences with their external actions and communication, meaning that improving authenticity heavily relies on the depth of self-reflection.
The scale consists of 12 items designed to measure the three factors: self-alienation, authentic living, and accepting external influence~\cite{barrett1998}.
Self-alienation highlights the inherent discrepancy between an individual's true experience and conscious awareness.
A divergence between the two suggests the individual feels disconnected or unfamiliar with their true self.
Authentic living assesses how well an individual's actions and emotions align with their internal state awareness.
Accepting external influence reflects an individual's propensity to be swayed by others, representing the impact of their social surroundings.
In the questionnaires for this scale, participants respond to each item using a seven-point Likert scale, ranging from 1 (``does not describe me at all'') to 5 (``describes me very well'').

\subsubsection{Behavioral Intention}

Behavioral intention was prepared to evaluate the clients' attitudes toward using the system.
The Technology Acceptance Model~\cite{Davis1989Perceived} guides the concept of behavioral intention, which explains users' attitudes towards technologies and is frequently used to evaluate how likely individuals are to use the technologies.
As we confirmed in the workshop (\secref{sec:workshop-findings}), clients' readiness is key to the outcome of their experience of text coaching.
Moreover, their intention to use the system depends on the quality of the experience.
Therefore, we thought this metric would provide useful context for analyzing the qualitative results of the semi-structured interviews.
We used the questionnaire from a previous study~\cite{Venkatesh2003User} to measure behavioral intention. This questionnaire consisted of three questions assessing respondents' intentions to continue using a system, with responses scored on a scale from 1 to 5.

\subsection{Procedure}

\begin{figure*}[t]
    \centering
    \includegraphics[width=0.75\textwidth]{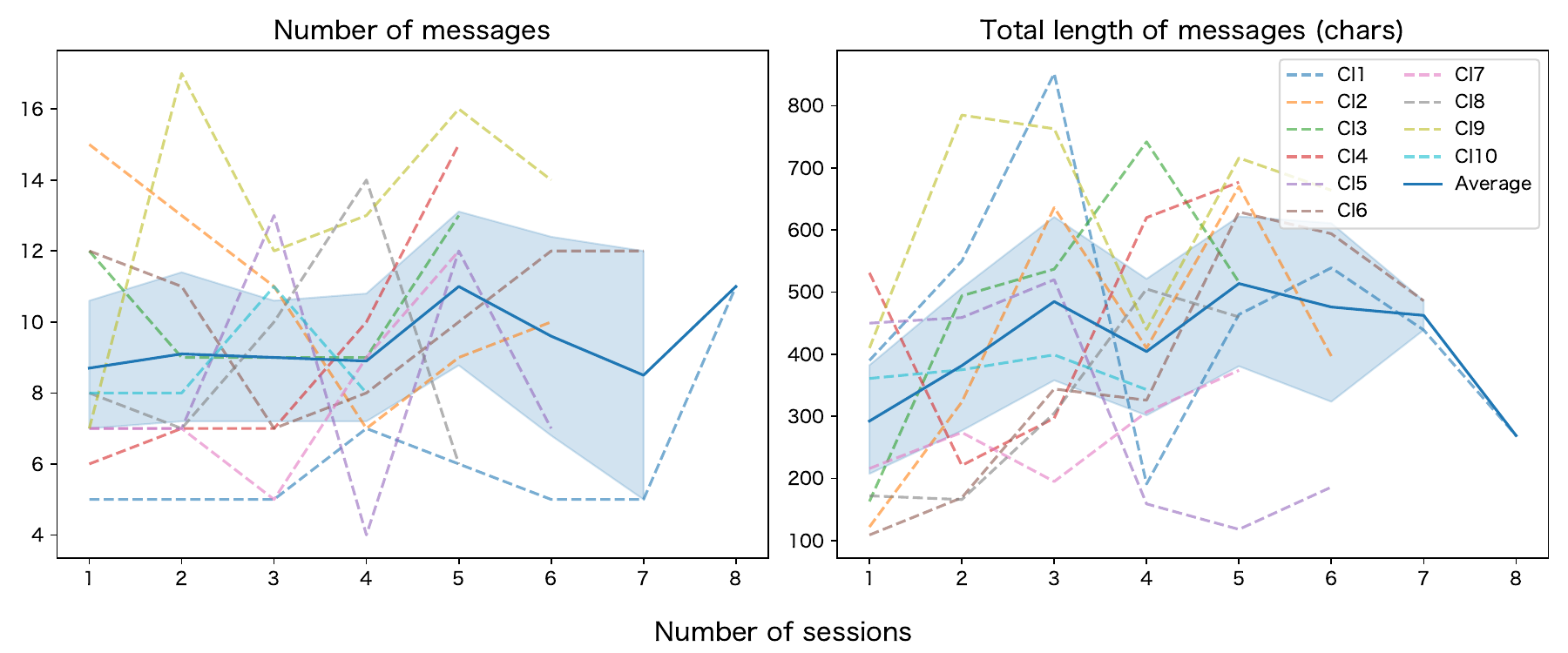}
    \caption{(Left) The number of messages each client sent to the chatbot coach per session. (Right) The total length of the messages per session in the number of characters. The blue area highlights indicate the 95\% confidence interval of the average value.}
    \Description{This figure shows two line charts. The left one presents the transition of the number of messages per session against the number of sessions each client has had. The right one presents the transition of the total length of the messages in the same manner. Each graph has ten lines corresponding to the clients, and a 95\% confidence interval of their average score is highlighted in blue.}
    \label{fig:usage}
\end{figure*}

\figref{fig:procedure} shows the procedure of our study.
First, an experimenter explained the prototype to the coaches by showing examples and having them play the system for a while.
Then, the coach conducted a face-to-face session with the client to explain the system's use and set the text coaching goal, which took roughly 30 minutes.
Specifically, they discussed their current issue, how they would utilize the chatbot, the goal state after the two weeks, and how often they planned to interact with the chatbot; this phase was informed by the workshop.
At the end of the session, the clients sent the answers to these questions on Google Form.
The answers were then used for configuring the system, as described in \secref{sec:method-prototype}.
This form also included questionnaires for behavioral intention.
Then, in the next two weeks, the client used the text-coaching system at their own pace; we did not set a dedicated time, nor did we remind them to do it, to preserve ecological validity. 
After each text coaching, they were navigated to answer questionnaires about the authenticity scale and behavioral intention as well as share any comments about the experience.
During the two-week trial period, there were no face-to-face sessions between the coach and the client.
After this period, the client and coach reflected on their text-coaching experience, discussing how they used the system and whether it met their goals.
Lastly, after the session, the experimenter conducted semi-structured interviews with the coach and the client individually, which took roughly 30 minutes.
For the coaches, they asked a series of questions: ``Compared to your usual coaching, how would you describe the coaching experience made possible by the AI technology?'', ``What are the advantages and disadvantages of chatbot coach?'', and ``Would you like to keep using the system in your coaching, and if yes, how?''.
On the other hand, the clients were asked about the following: ``How did your behavior change during the coaching period?'', ``Were you able to achieve the goal you had set in the beginning?'', ``How would you change your use of the system if you did it again?'', ``How has your trust in the chatbot coach changed over time?'', and ``Would you like to keep using the system, and if yes, how?''.

\section{Results}
\label{sec:results}

\subsection{Chat Behavior}
\label{sec:results-behavior}

We first examined how the clients used the chatbot coach (See~\figref{fig:usage}). 
During the two-week trial period, the average number of text-coaching sessions was 5.7 ($\text{SD}=1.1$).
The variance is understandable, given the expected number had been arbitrarily set during the first face-to-face session.
Note that, although we did not remind them, all participants except one client met the declared number.
Additionally, while there was a variance in the chat behavior by participants and sessions, their average number of messages and their lengths were sustained over the sessions.
The results suggested that the participants engaged in text coaching throughout the study.

\begin{figure*}[t]
    \centering
    \includegraphics[width=\textwidth]{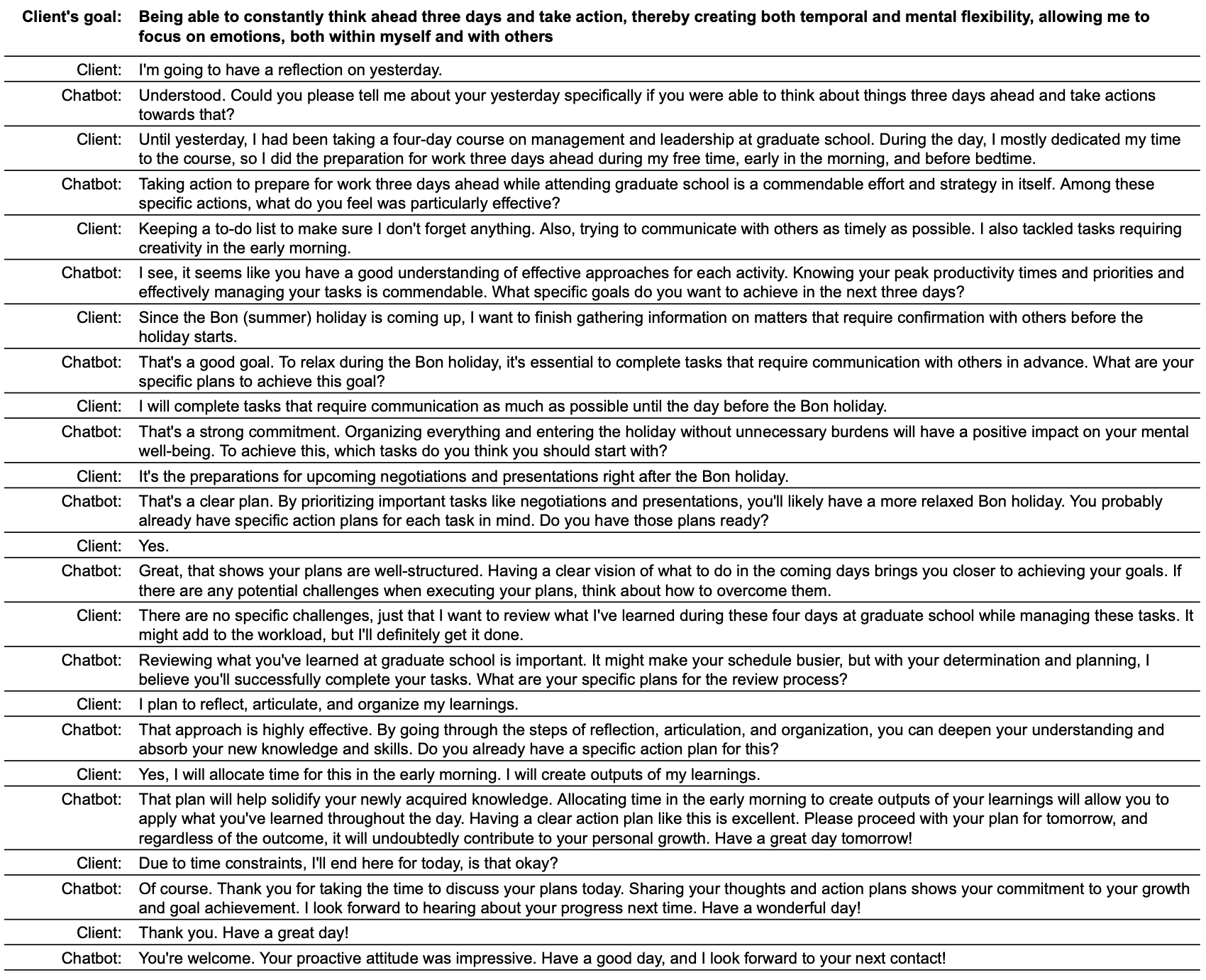}
    \caption{Example from one client's messages with the chatbot coach. It is observed that the chatbot often acknowledges the client's actions and asks questions that can further break down the problems they face.}
    \Description{The image of an example dialogue between a client and the chatbot coach. The first line shows the clients' goal, which had been decided in a session with the human coach.}
    \label{fig:example}
\end{figure*}

\figref{fig:example} provides an actual chat history of one client, which was presented with their permission.
This shows that the client promoted her reflection within 10--15 messages.
Also, we can infer that the responses from the chatbot coach motivated the client to take action towards her goal, as the chatbot effectively utilized the coaching skill of ``acknowledgment''~\cite{Whitworth1998}.
Specifically, coaches are known to induce the client's new behavior by acknowledging a client's action and, more importantly, reflecting on how that action honored their values.
To understand their experience of using the chatbot coach and inform the answer for~\textbf{RQ3}, we analyzed their comments during our semi-structured interviews in the next section.

\subsection{Semi-Structured Interviews}

The participants' (\ie both coaches' and clients') responses in the semi-structured interviews were analyzed using open coding~\cite{StrCor90}.
Through iterative refinement processes, we obtained four topics, as follows.

\subsubsection{Power of a Chatbot Coach as a Promoter of Clients' Actions When Human Coach Is Not Present}
\label{sec:results-semi-power}

Firstly, we found that all clients responded affirmatively about the power of the chatbot in promoting their actions, like:
\begin{quote}
    Being able to communicate with the coach at my preferred time was casual and nice. This was helpful in maintaining my motivation. (Cl8)
\end{quote}
Similar comments were also obtained from the coach side, such as:
\begin{quote}
    It was quite convenient that the coaching proceeded without the trouble of fixing the session schedule in advance. (Co6)
\end{quote}
This confirms the discussions we observed in the workshop.
Moreover, the clients' comments shed light on how their reflection was prompted through the dialogue,
\begin{quote}
    By being repeatedly and persistently asked, like ``then, what exactly will the next step you do?'', I was able to clarify what kind of preparation and planning was necessary. (Cl2)
\end{quote}
This advantage that the chatbot can help clients break down actions underpins our prompt design guided by the workshop.
We believe this has been made possible by the LLM's zero-shot reasoning capability in diverse problem-solving tasks~\cite{DBLP:conf/nips/KojimaGRMI22}.

In addition, some clients sympathized with the chatbot more than we had expected.
\begin{quote}
    Even receiving simple replies like ``that's good'' made me happy, even though I knew the opponent was an AI. I found it surprisingly delightful, and it helped me make progress. (Cl1)
\end{quote}
This point helped the clients take practical action based on their reflection, confirming the effect of the chatbot's acknowledgment that we discussed in \secref{sec:results-behavior}.
These comments suggest the power of LLM-powered chatbots in inducing clients' reflection and sustainable engagement for behavioral changes.

\subsubsection{Limitation of a Chatbot Coach in Inducing Deep Reflection}
\label{sec:results-semi-limitation}

At the same time, the specific aspect of the chatbot that can be derived from the nature of LLMs appeared to be a limitation.
\begin{quote}
    As the conversation continued, I found it would be beneficial if the chatbot asked in-depth questions like ``Isn't the goal you set initially a bit lenient?'' or ``What's the true significance of pursuing this goal?'' (Cl7)
\end{quote}
\begin{quote}
    Questions like identifying behaviors the client unconsciously avoids would deepen the conversation but were not observed. I believe that such aspects would not been covered unless I proactively intervened in the conversation as a human's role. (Co2)
\end{quote}
This would be intrinsic to the LLM we used, GPT-4, which is trained to follow users' intention~\cite{DBLP:conf/nips/Ouyang0JAWMZASR22} so that it would not generate exceptionable responses.
On the other hand, to induce deep reflection of clients, coaching sometimes requires questions that challenge the clients and would make them uncomfortable~\cite{Whitworth1998}.
This point can be a limitation of chatbot coaches, conversely suggesting the necessity of human coaches' involvement in the process.

We also would like to note that the button to ask human coaches for guidance during the text coaching sessions (\figref{fig:prototype}F) was rarely used.
In this regard, one client commented:
\begin{quote}
    I did not have much opportunity to use the button because I was satisfied with the fact that some actions were progressing even without deep, challenging questions. (Cl8)
\end{quote}
One coach also mentioned:
\begin{quote}
    Estimating the timing to interject a sharp retort or comment is very difficult even for a human coach, and it's very important. Personally, I think there is a need to guide the conversation from our end, rather than through text, but through face-to-face conversation. (Co10)
\end{quote}
This would imply the need for human coaches to carefully monitor the communication between clients and the chatbot coach, as initiating such communication from the client side would be challenging.

\subsubsection{Importance of Clarifying Goals and Securing Commitment with a Human Coach}

Despite such limitations, the clients favored the overall experience of the text coaching, maintaining the engagement during the trial as we observed in~\secref{sec:results-behavior}.
Their comments revealed that this was due to the designed approach of blending human and chatbot coaches, even though they were not informed about the discussion results of the previous workshop.
\begin{quote}
    Making an initial commitment with the human coach regarding the extent of using the chatbot became the motivation for actually following through with it. (Cl4)
\end{quote}
\begin{quote}
    The fact that the human coach was keeping track of my progress positively motivated me to engage in text coaching with the chatbot. (Cl3)
\end{quote}
A similar perspective was provided also by the coaches.
\begin{quote}
    Once we set a goal and made a promise about the usage with solid motivation, the entire process proceeded without any effort. This would be one of the optimal ways to introduce the chatbot to executive coaching. (Co6)
\end{quote}
These comments confirm the effectiveness of our design in promoting the blended initiatives of the human and chatbot coaches, which was informed by the workshop.

Moreover, we found that the blended initiative can foster the self-disclosure of the clients.
\begin{quote}
    I can flatly talk to the chatbot. With a human coach, it takes time to realize trust, and sometimes it's a little difficult to be honest. The chatbot was an easy contact of communication between the human coach and me. (Cl9)
\end{quote}
\begin{quote}
    Some people need a long time to disclose themselves to coaches. But, since the chatbot can provide frequent communication, their disclosure can be accelerated. I learned that the frequency of the text coaching sessions can be a key to successful outcomes. (Co4)
\end{quote}
These comments were particularly interesting because it was the opposite of the expectation of the literature~\cite{Terblanche2020Factors} that suggested the difficulty of clients in trusting AIs.
We suppose this would not be realized without the human coaches who secure the commitment of the clients' frequent chatbot use.

\subsubsection{Direction for the Improved Blending Coaching}

Given such observations and limitations, some coaches suggested directions for enabling better collaboration between clients and chatbot coaches.
\begin{quote}
    It would be good to allocate 60 minutes for the first session before starting to use the chatbot coach to confirm the significance of the goal a client sets and to adjust the difficulty level of the goal. This would reduce the risk of clients working on goals that are not essential or too easy. (Co4)
\end{quote}
\begin{quote}
    When the client did not seem to be making good progress, it would have been better to have an opportunity to talk for five minutes or so instead of communicating via text. When I was writing my reply to the client in text, I thought this text could also be generated by an AI with some tuning. Conversely, I felt that the value of a human coach lies in the ability to ask tough questions in person, taking into consideration the subtleties in clients' speech and other aspects. (Co8)
\end{quote}
We believe that the values of our study are not limited to confirming the effectiveness of our design but include these practical insights provided by coaches who actually experienced the blended approach.
Later in \secref{sec:disc}, we summarize our findings to foster the effective use of LLM-powered chatbots in executive coaching.

\begin{figure}[t]
    \centering
    \includegraphics[width=\columnwidth]{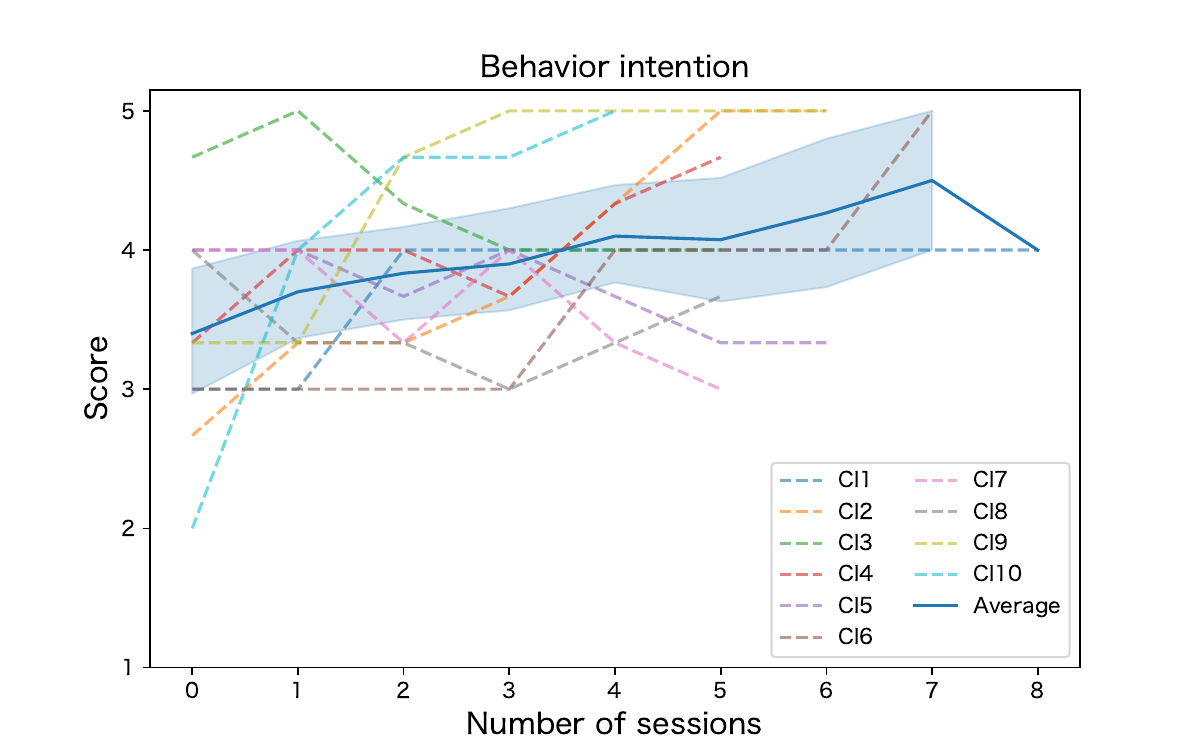}
    \caption{The transition of the score of the participated clients' behavioral intention to use the chatbot coach. The blue area highlights the 95\% confidence interval of the average score.}
    \Description{This figure shows the transition of the behavioral intention score responded to by the clients in a line chat. There are ten lines in the graph corresponding to the clients, and a 95\% confidence interval of their average score is highlighted in blue.}
    \label{fig:intention}
\end{figure}

\subsection{Clients' Authenticity Scale and Behavioral Intention}

\begin{figure*}[t]
    \centering
    \includegraphics[width=\textwidth]{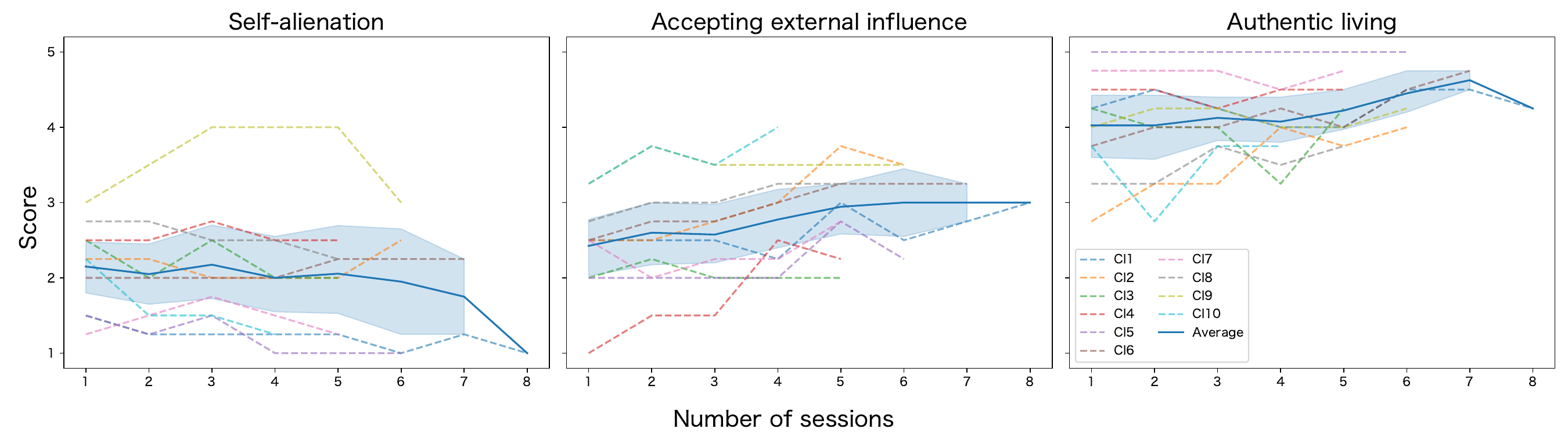}
    \caption{The transitions of the scores of the three factors of the participated clients' authenticity. The blue area highlights the 95\% confidence interval of the average score.}
    \Description{This figure shows three line chats, each of which presents the score's transition for the authenticity scale's three factors: from left to right, self-alienation, accepting external influence, and authentic living. The x-axis is the number of sessions, while the y-axis shows the score. Each graph has ten lines corresponding to the clients, and a 95\% confidence interval of their average score is highlighted in blue.}
    \label{fig:authenticity}
\end{figure*}

We also analyzed the clients' evaluations of their authenticity and behavioral intention, which were collected after the end of each text coaching session.
\figref{fig:intention} shows the transition of their scores of the behavioral intention, in which we did not find negative responses once they started using the chatbot coach.
Rather, we found many participants who increased their scores as they had more sessions, except for two clients.
We inferred that their consistent use of the chatbot, as discussed in ~\secref{sec:results-behavior}, resulted from their positive acceptance.
At the same time, we confirmed that the two clients who showed a decreasing trend mentioned the limitations of the chatbot during the semi-structured interviews, as described in \secref{sec:results-semi-limitation}.
This conversely implied the importance of human coaches' involvement; clients' behavioral intentions might decay without such involvement, especially when they need deeper reflection.

\figref{fig:authenticity} shows the transitions of the clients' authenticity scores.
We observed the most dynamic transition regarding accepting external influence, reflecting that the clients started to change their behavior based on the conversations with the chatbot coach.
Given that such actions functioned toward achieving their declared goals (see \secref{sec:results-semi-power}), this suggests the effectiveness of the chatbot coach in inducing positive outcomes through text coaching.
Also, we can see a decreasing trend in self-alienation.
Though it was not so significant, we conjectured that reflection through dialogue with the chatbot coach could enhance their conscious awareness.
At the same time, given the observed limitations of the chatbot (see \secref{sec:results-semi-limitation}), the effect could be further enhanced by exploring the optimal blending of human and chatbot coaches, for instance, as suggested by Co8 as an improvement direction.

In conclusion, our findings for \textbf{RQ3} revealed that the chatbot coach significantly contributed to fostering clients' reflection and inspiring their actions, especially in the absence of a human coach.
Furthermore, the integrated design of human and chatbot coaches, as derived from insights in \textbf{RQ2}, proved effective, encouraging consistent client engagement with the chatbot.
The limitations identified, inherent to current LLMs, underscored the necessity of this blended approach, which enables clients to have deep self-reflection effectively.
The outcomes associated with \textbf{RQ3} highlight the potential to broaden the reach of executive coaching, making it accessible to all those in need through the support of LLMs.

\section{Discussion}
\label{sec:disc}

Through building a prototype with insights from coaches and conducting an empirical study in actual coaching scenes, we have shed light on the plausible form of using LLM-powered chatbots effectively for one's leadership growth.
Lastly, we discuss the study's implications for the HCI community and derive a guideline for deploying such chatbots in HRD practice.

\subsection{Types of Reflections Chatbot Coaches Support}

The results of the semi-structured interviews suggested that the LLM-powered chatbot coach facilitated clients' reflection while uncovering their limitations.
This ties into the discussion on the contrasting learning concepts, single-loop learning and double-loop learning~\cite{argyris1977double}.
Single-loop learning involves adjusting and correcting existing frameworks and policies to address discrepancies without changing underlying assumptions or values.
In contrast, double-loop learning goes further by analyzing and potentially altering the fundamental assumptions or governing values, facilitating more profound organizational changes and improvements.
This scheme clarifies the advantages and disadvantages of introducing the LLM-powered chatbot coach.
First, our results suggest that the chatbot coach can support clients' single-loop learning by iteratively asking questions to clarify the steps toward the set goal while acknowledging their actions, as shown in~\figref{fig:example}.
In our study, the human coach and the client favored the benefit of chatbots always being available.
This results not only in reducing the workload of human coaches but also in fostering the reflection of clients on their convenient occasions.
However, as discussed in~\secref{sec:results-semi-limitation}, it is challenging to facilitate their double-loop learning because posing incisive and pushing questions can be regarded as too sensitive for the chatbot coach.
Specifically, we need to acknowledge that making an LLM-powered chatbot capable of asking challenging questions increases the risk of manipulating clients' emotions in a harmful manner, as Cabrera~\etal~\cite{DBLP:conf/iwbbio/CabreraLMR23} discussed in the context of why chatbots need human supervision.
Therefore, while Liu~\etal~\cite{Liu2021} reported the effectiveness of introducing the idea of double-loop learning into skill training, it is also suggested that automating such learning processes is sometimes difficult, especially when deep reflection is required.

\begin{figure*}[t]
    \begin{center}
        \includegraphics[width=0.8\textwidth]{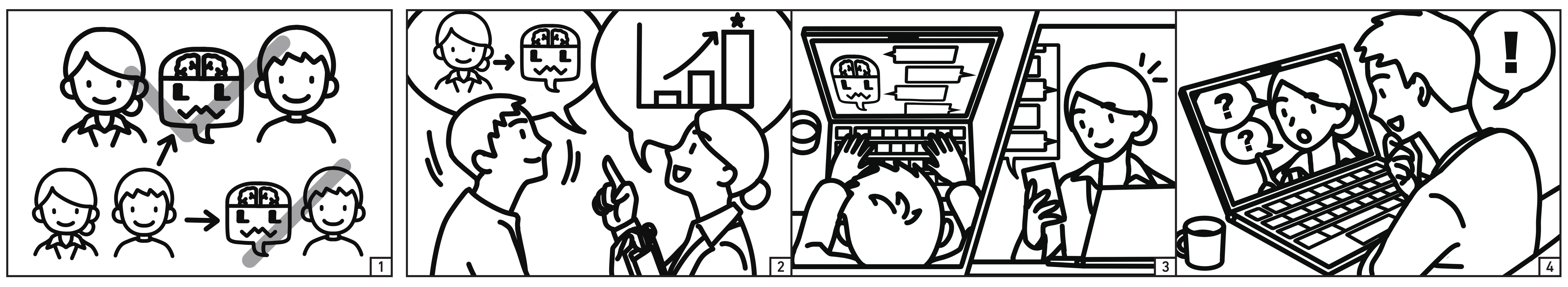}
    \end{center}
    \caption{Guideline for the blended approach of human and LLM-powered chatbot coach for leadership growth in executive coaching, synthesized through our study. The human coach is suggested to follow the steps to introduce the chatbot coach successfully and to augment clients' reflection: 1) introducing the chatbot as complementary text coaching, 2) conducting pre-session to foster a client's readiness before initiating chatbot-driven coaching, 3) maintaining a communication channel with the client during the text coaching period, and 4) monitoring client progress and identify moments to intervene.}
    \Description{This figure denotes four points of the guideline. 1) Introduce chatbot as complementary text coaching---The icons of coach, client, LLM are used to illustrate this. 2) Conduct a pre-session before introducing the chatbot coach – The conversation in which the coach explains the scheme is illustrated. 3) Maintain communication channel after text coaching---Two scenes of the client using the chatbot on a laptop and the coach receiving a notification about the progress on a smartphone are illustrated. 4) Monitor client progress and identify moments to intervene critically – A video conversation between the coach and client is illustrated.}
    \label{fig:teaser}
\end{figure*}

\subsection{Roles of Human Coach in Executive Coaching Dialogue}

Our study also revealed the advantages and disadvantages of human coaches, especially through the lens of single/double-loop learning.
For example, we can infer from the obtained comments that the human coaches were crucial for inducing clients' deep reflection through potentially uncomfortable inquiry.
Prior work suggested that professional coaches can interpret subtle signals of clients, such as nonverbal behavioral cues during face-to-face conversation, to pose such questions~\cite{DBLP:conf/chi/ArakawaY19}.
Our study also suggested that the existence of human coaches can contribute to clients maintaining their engagement with the entire coaching experience. 
However, human coaches are not always available, which limits the opportunity for clients to have reflection instantly; as discussed in \secref{sec:bg}, the industry of executive coaching faces the problem of lacking proficient coaches.
Together with the advantages and disadvantages of chatbot coaches, the effective blending of both parties is demanded, in contrast to focusing on the engagement with one side, and our study also uncovered multiple factors key to achieving it, which we summarize in the next section.

\subsection{Toward Deployment of Chatbot Coach for Leadership Growth}
\label{sec:disc-toward}

To conclude this paper, we summarize the guided flow for human coaches to follow to blend human and chatbot coaches in executive coaching effectively, the overview of which is presented in~\figref{fig:teaser}.

\subsubsection{Introduce Chatbot as Complementary Text Coaching}
\label{sec:disc-toward-1}

The workshop indicated that delegating all components human coaches take in executive coaching to chatbots is not reasonable, given the significance of actual human presence in influencing clients' behavior. 
On the other hand, our study confirmed chatbots' effectiveness in reducing the workload of text coaching to keep the clients' attitude to improve their behavior.
Thus, clarifying the point that chatbots serve as a complement to traditional face-to-face sessions is important.

\subsubsection{Conduct Pre-Session Before Introducing Chatbot Coach}
\label{sec:disc-toward-2}

Our study also suggested the importance of setting a clear goal for text coaching to ensure a successful experience.
In detail, due to the nature of chatbots, efficient usage naturally forms within the client through use when their readiness is enough.
Thus, the human coach is encouraged to spend a certain amount of time with the client to adjust the difficulty level of the goal and expected outcome, which would last 30--60 minutes.

\subsubsection{Maintain Communication Channel after Text Coaching}
\label{sec:disc-toward-3}

Furthermore, the presence of the human coach should be conveyed during the text coaching period to keep the client's motivation and readiness.
A communication channel after text coaching, \eg the client sending its summary to the human coach and the coach leaving reactions to it, would be recommended.

\subsubsection{Monitor Client Progress and Identify Moments to Intervene Critically}
\label{sec:disc-toward-4}

Our study confirmed that the intervention from the human coach is desired to induce deeper reflection, \eg double-loop learning.
Currently, the best way to have such opportunities remains an open question.
As implied in~\secref{sec:results-semi-limitation}, it would be hard for clients themselves to initiate such conversation as it is by nature mentally challenging.
Therefore, the coach needs to monitor their progress by, for instance, paying attention to the summary report the clients send or introducing a quick survey after the text coaching to ask about their satisfaction.

\subsection{Implication for Chatbot-Based Reflection Support in Other Settings}

The advantage of the chatbot coach we discussed in~\secref{sec:results-semi-power} shares aspects with that of the chatbots in other domains as discussed in~\secref{sec:bg-chatbot}, that is, being available when the human coach is not present and providing consistent feedback or answers.
Furthermore, the recent advancement in LLM enables chatbots to ask natural and reasonable questions to foster actions to achieve the planned goal, showing a greater capability than a previously-explored chatbot, \eg scripted feedback of the health coach in~\cite{DBLP:journals/pacmhci/MitchellMCTDSM21}.
Meanwhile, in the context of executive coaching, the importance of nurturing a space for critical introspection has been highlighted, and we conclude that a blend of human and chatbot coaches is necessary.
Such a design can apply to reflection support in other domains, where users need to be rigorously challenged by external sources to transition into a more ideal state.
For example, Li~\etal~\cite{DBLP:conf/cui/LiLLLL23} explored the design of LLM-powered chatbots to encourage moderate smartphone use and emphasized the importance of personalized dialogue.
This work can be augmented by our exploration into a complementary strategy, namely, the blended form of human and chatbot coaches.
In addition, Lee~\etal~\cite{DBLP:conf/chi/LeeAACLI19} showed that, for mental health care, not only providing care from a chatbot but also asking users to provide care for a (vulnerably conditioned) chatbot can be effective.
This practice can be used in a blended manner, in which human therapists contribute to clarifying the goal of the interaction with a chatbot and maintaining the motivation to continue caring for the chatbot.
Our empirical findings highlight an emerging theme that demands this blended approach, especially in the era of rapidly evolving LLMs.

\section{Conclusion and Future Work}

We clarified the potential of an LLM-powered chatbot to foster one's deep reflection to achieve leadership growth by taking executive coaching as a study field.
Our studies in actual coaching scenes revealed the strength of both human coach and chatbot coach and synthesized a way to combine them to realize optimal reflection experience.
The current study involves a limited number of coaching organizations, which might result in homogeneous samples sharing a single cultural context.
Thus, more participants (especially coaches) at different organizations are expected to verify the findings further.

In addition, since individual coaching experience varies significantly, conducting a comparison study with a baseline condition was difficult without involving a large participant pool.
Rather, we co-designed with both coaches and clients to advance the practice with an emerging technology.
Again, a larger user study with a baseline condition (\eg dropping one of the key steps we identified in \secref{sec:disc-toward}) will bring additional evidence to quantify the effect.
It would be beneficial if we could conduct the study in the long term because it would reduce the possibility that a novelty effect occurred, as pointed out by Weber~\etal~\cite{DBLP:journals/icom/WeberMAJLW23}.

Discussion on ethical considerations is also crucial to push the deployment of the chatbot forward.
For example, clarifying the role of the chatbot and the scope of information sharing when introducing the chatbot (see \secref{sec:disc-toward-1}) would be crucial in terms of both data privacy and psychological safety.
Here, the standard in executive coaching about confidentiality~\cite{Greenfield2004} can also be applied to this blended approach.
At the same time, the necessity of monitoring the risk of introducing bias or discrimination in chatbots~\cite{DBLP:conf/conversations/FeineGMM19} is common to us, while we believe that our design of trying not to personify the chatbot can be a remedy.

Furthermore, we can extend the blended approach to develop a socio-technical tool that allows practitioners to leverage the approach in other domains, as Sadek~\etal~\cite{DBLP:conf/cui/SadekCM23} suggested.
At the same time, we need to be aware that the outcome of chatbot coaches may evolve with the future advancement of LLMs.
Nevertheless, we believe that our qualitative data show the enduring value of human coaches and consistently support the benefits of the blended approach.

\begin{acks}
This work was supported in part by JST ACT-X Grant Number JPMJAX200R and JSPS KAKENHI Grant Numbers JP21J20353. Also, this work was conducted in collaboration with and under the supervision of Teambox, Inc.
\end{acks}

\bibliographystyle{ACM-Reference-Format}
\bibliography{paper}


\begin{thebibliography}{69}


\ifx \showCODEN    \undefined \def \showCODEN     #1{\unskip}     \fi
\ifx \showDOI      \undefined \def \showDOI       #1{#1}\fi
\ifx \showISBNx    \undefined \def \showISBNx     #1{\unskip}     \fi
\ifx \showISBNxiii \undefined \def \showISBNxiii  #1{\unskip}     \fi
\ifx \showISSN     \undefined \def \showISSN      #1{\unskip}     \fi
\ifx \showLCCN     \undefined \def \showLCCN      #1{\unskip}     \fi
\ifx \shownote     \undefined \def \shownote      #1{#1}          \fi
\ifx \showarticletitle \undefined \def \showarticletitle #1{#1}   \fi
\ifx \showURL      \undefined \def \showURL       {\relax}        \fi
\providecommand\bibfield[2]{#2}
\providecommand\bibinfo[2]{#2}
\providecommand\natexlab[1]{#1}
\providecommand\showeprint[2][]{arXiv:#2}

\bibitem[\protect\citeauthoryear{Alqahtani, Jay, and Vigo}{Alqahtani
  et~al\mbox{.}}{2020}]%
        {DBLP:journals/imwut/AlqahtaniJV20}
\bibfield{author}{\bibinfo{person}{Deemah~A. Alqahtani},
  \bibinfo{person}{Caroline Jay}, {and} \bibinfo{person}{Markel Vigo}.}
  \bibinfo{year}{2020}\natexlab{}.
\newblock \showarticletitle{The Effect of Goal Moderation on the Achievement
  and Satisfaction of Physical Activity Goals}.
\newblock \bibinfo{journal}{\emph{Proc. {ACM} Interact. Mob. Wearable
  Ubiquitous Technol.}} \bibinfo{volume}{4}, \bibinfo{number}{4}
  (\bibinfo{year}{2020}), \bibinfo{pages}{116:1--116:18}.
\newblock
\urldef\tempurl%
\url{https://doi.org/10.1145/3432209}
\showDOI{\tempurl}


\bibitem[\protect\citeauthoryear{Amershi, Weld, Vorvoreanu, Fourney, Nushi,
  Collisson, Suh, Iqbal, Bennett, Inkpen, Teevan, Kikin{-}Gil, and
  Horvitz}{Amershi et~al\mbox{.}}{2019}]%
        {DBLP:conf/chi/AmershiWVFNCSIB19}
\bibfield{author}{\bibinfo{person}{Saleema Amershi}, \bibinfo{person}{Daniel~S.
  Weld}, \bibinfo{person}{Mihaela Vorvoreanu}, \bibinfo{person}{Adam Fourney},
  \bibinfo{person}{Besmira Nushi}, \bibinfo{person}{Penny Collisson},
  \bibinfo{person}{Jina Suh}, \bibinfo{person}{Shamsi~T. Iqbal},
  \bibinfo{person}{Paul~N. Bennett}, \bibinfo{person}{Kori Inkpen},
  \bibinfo{person}{Jaime Teevan}, \bibinfo{person}{Ruth Kikin{-}Gil}, {and}
  \bibinfo{person}{Eric Horvitz}.} \bibinfo{year}{2019}\natexlab{}.
\newblock \showarticletitle{Guidelines for human-AI interaction}. In
  \bibinfo{booktitle}{\emph{Proceedings of the 2019 {ACM} {CHI} Conference on
  Human Factors in Computing Systems}}. \bibinfo{publisher}{{ACM}},
  \bibinfo{address}{New York, NY}, \bibinfo{pages}{3}.
\newblock
\urldef\tempurl%
\url{https://doi.org/10.1145/3290605.3300233}
\showDOI{\tempurl}


\bibitem[\protect\citeauthoryear{Arakawa, Maeda, and Yakura}{Arakawa
  et~al\mbox{.}}{2024}]%
        {DBLP:journals/corr/abs-2402-11145}
\bibfield{author}{\bibinfo{person}{Riku Arakawa}, \bibinfo{person}{Kiyosu
  Maeda}, {and} \bibinfo{person}{Hiromu Yakura}.}
  \bibinfo{year}{2024}\natexlab{}.
\newblock \showarticletitle{Supporting Experts with a Multimodal
  Machine-Learning-Based Tool for Human Behavior Analysis of Conversational
  Videos}.
\newblock \bibinfo{journal}{\emph{CoRR}}  \bibinfo{volume}{abs/2402.11145}
  (\bibinfo{year}{2024}).
\newblock
\urldef\tempurl%
\url{https://doi.org/10.48550/ARXIV.2402.11145}
\showDOI{\tempurl}


\bibitem[\protect\citeauthoryear{Arakawa and Yakura}{Arakawa and
  Yakura}{2019}]%
        {DBLP:conf/chi/ArakawaY19}
\bibfield{author}{\bibinfo{person}{Riku Arakawa} {and} \bibinfo{person}{Hiromu
  Yakura}.} \bibinfo{year}{2019}\natexlab{}.
\newblock \showarticletitle{REsCUE: {A} framework for REal-time feedback on
  behavioral CUEs using multimodal anomaly detection}. In
  \bibinfo{booktitle}{\emph{Proceedings of the 2019 {ACM} {CHI} Conference on
  Human Factors in Computing Systems}}. \bibinfo{publisher}{{ACM}},
  \bibinfo{address}{New York, NY}, \bibinfo{pages}{572}.
\newblock
\urldef\tempurl%
\url{https://doi.org/10.1145/3290605.3300802}
\showDOI{\tempurl}


\bibitem[\protect\citeauthoryear{Arakawa and Yakura}{Arakawa and
  Yakura}{2020}]%
        {DBLP:conf/chi/ArakawaY20}
\bibfield{author}{\bibinfo{person}{Riku Arakawa} {and} \bibinfo{person}{Hiromu
  Yakura}.} \bibinfo{year}{2020}\natexlab{}.
\newblock \showarticletitle{{INWARD:} {A} computer-supported tool for
  video-reflection improves efficiency and effectiveness in executive
  coaching}. In \bibinfo{booktitle}{\emph{Proceedings of the 2020 {ACM} {CHI}
  Conference on Human Factors in Computing Systems}}.
  \bibinfo{publisher}{{ACM}}, \bibinfo{address}{New York, NY},
  \bibinfo{pages}{1--13}.
\newblock
\urldef\tempurl%
\url{https://doi.org/10.1145/3313831.3376703}
\showDOI{\tempurl}


\bibitem[\protect\citeauthoryear{Argyris}{Argyris}{1977}]%
        {argyris1977double}
\bibfield{author}{\bibinfo{person}{Chris Argyris}.}
  \bibinfo{year}{1977}\natexlab{}.
\newblock \showarticletitle{Double loop learning in organizations}.
\newblock \bibinfo{journal}{\emph{Harvard business review}}
  \bibinfo{volume}{55}, \bibinfo{number}{5} (\bibinfo{year}{1977}),
  \bibinfo{pages}{115--125}.
\newblock


\bibitem[\protect\citeauthoryear{Barrett-Lennard}{Barrett-Lennard}{1999}]%
        {barrett1998}
\bibfield{author}{\bibinfo{person}{Godfrey~T Barrett-Lennard}.}
  \bibinfo{year}{1999}\natexlab{}.
\newblock \bibinfo{booktitle}{\emph{Carl Rogers' helping system: Journey and
  substance}}.
\newblock \bibinfo{publisher}{SAGE Publications}, \bibinfo{address}{London,
  UK}.
\newblock


\bibitem[\protect\citeauthoryear{Baumer, Khovanskaya, Matthews, Reynolds,
  Sosik, and Gay}{Baumer et~al\mbox{.}}{2014}]%
        {DBLP:conf/ACMdis/BaumerKMRSG14}
\bibfield{author}{\bibinfo{person}{Eric P.~S. Baumer}, \bibinfo{person}{Vera~D.
  Khovanskaya}, \bibinfo{person}{Mark Matthews}, \bibinfo{person}{Lindsay
  Reynolds}, \bibinfo{person}{Victoria~Schwanda Sosik}, {and}
  \bibinfo{person}{Geri Gay}.} \bibinfo{year}{2014}\natexlab{}.
\newblock \showarticletitle{Reviewing reflection: On the use of reflection in
  interactive system design}. In \bibinfo{booktitle}{\emph{Proceedings of the
  2014 {ACM} Designing Interactive Systems Conference}}.
  \bibinfo{publisher}{{ACM}}, \bibinfo{address}{New York, NY},
  \bibinfo{pages}{93--102}.
\newblock
\urldef\tempurl%
\url{https://doi.org/10.1145/2598510.2598598}
\showDOI{\tempurl}


\bibitem[\protect\citeauthoryear{Beaudry, Consigli, Clark, and
  Robinson}{Beaudry et~al\mbox{.}}{2019}]%
        {Beaudry2019Getting}
\bibfield{author}{\bibinfo{person}{Jeremy Beaudry}, \bibinfo{person}{Alyssa
  Consigli}, \bibinfo{person}{Colleen Clark}, {and} \bibinfo{person}{Keith~J.
  Robinson}.} \bibinfo{year}{2019}\natexlab{}.
\newblock \showarticletitle{Getting ready for adult healthcare: Designing a
  chatbot to coach adolescents with special health needs through the
  transitions of care}.
\newblock \bibinfo{journal}{\emph{Journal of Pediatric Nursing}}
  \bibinfo{volume}{49} (\bibinfo{year}{2019}), \bibinfo{pages}{85--91}.
\newblock
\urldef\tempurl%
\url{https://doi.org/10.1016/j.pedn.2019.09.004}
\showDOI{\tempurl}


\bibitem[\protect\citeauthoryear{Benke, Vetter, and Maedche}{Benke
  et~al\mbox{.}}{2021}]%
        {DBLP:conf/cscw/BenkeVM21}
\bibfield{author}{\bibinfo{person}{Ivo Benke}, \bibinfo{person}{Sebastian
  Vetter}, {and} \bibinfo{person}{Alexander Maedche}.}
  \bibinfo{year}{2021}\natexlab{}.
\newblock \showarticletitle{LeadBoSki: {A} Smart Personal Assistant for
  Leadership Support in Video-Meetings}. In \bibinfo{booktitle}{\emph{Companion
  Publication of the 2021 {ACM} Conference on Computer Supported Cooperative
  Work and Social Computing}}. \bibinfo{publisher}{{ACM}},
  \bibinfo{pages}{19--22}.
\newblock
\urldef\tempurl%
\url{https://doi.org/10.1145/3462204.3481764}
\showDOI{\tempurl}


\bibitem[\protect\citeauthoryear{Bentvelzen, Wozniak, Herbes, Stefanidi, and
  Niess}{Bentvelzen et~al\mbox{.}}{2022}]%
        {DBLP:journals/imwut/BentvelzenWHSN22}
\bibfield{author}{\bibinfo{person}{Marit Bentvelzen}, \bibinfo{person}{Pawel~W.
  Wozniak}, \bibinfo{person}{Pia S.~F. Herbes}, \bibinfo{person}{Evropi
  Stefanidi}, {and} \bibinfo{person}{Jasmin Niess}.}
  \bibinfo{year}{2022}\natexlab{}.
\newblock \showarticletitle{Revisiting reflection in {HCI:} Four design
  resources for technologies that support reflection}.
\newblock \bibinfo{journal}{\emph{Proceedings of the {ACM} on Interactive,
  Mobile, Wearable and Ubiquitous Technologies}} \bibinfo{volume}{6},
  \bibinfo{number}{1} (\bibinfo{year}{2022}), \bibinfo{pages}{2:1--2:27}.
\newblock
\urldef\tempurl%
\url{https://doi.org/10.1145/3517233}
\showDOI{\tempurl}


\bibitem[\protect\citeauthoryear{Bridgeman and Giraldez-Hayes}{Bridgeman and
  Giraldez-Hayes}{2023}]%
        {Bridgeman2023Using}
\bibfield{author}{\bibinfo{person}{James Bridgeman} {and}
  \bibinfo{person}{Andrea Giraldez-Hayes}.} \bibinfo{year}{2023}\natexlab{}.
\newblock \showarticletitle{Using artificial intelligence-enhanced video
  feedback for reflective practice in coach development: Benefits and potential
  drawbacks}.
\newblock \bibinfo{journal}{\emph{Coaching: An International Journal of Theory,
  Research and Practice}} (\bibinfo{year}{2023}), \bibinfo{pages}{1--18}.
\newblock
\urldef\tempurl%
\url{https://doi.org/10.1080/17521882.2023.2228416}
\showDOI{\tempurl}


\bibitem[\protect\citeauthoryear{Cabrera, Loyola, Maga{\~{n}}a, and
  Rojas}{Cabrera et~al\mbox{.}}{2023}]%
        {DBLP:conf/iwbbio/CabreraLMR23}
\bibfield{author}{\bibinfo{person}{Johana Cabrera},
  \bibinfo{person}{Mar{\'{\i}}a~Soledad Loyola}, \bibinfo{person}{Irene
  Maga{\~{n}}a}, {and} \bibinfo{person}{Rodrigo Rojas}.}
  \bibinfo{year}{2023}\natexlab{}.
\newblock \showarticletitle{Ethical Dilemmas, Mental Health, Artificial
  Intelligence, and LLM-Based Chatbots}. In
  \bibinfo{booktitle}{\emph{Proceedings of the 10th International
  Work-Conference on Bioinformatics and Biomedical Engineering}},
  Vol.~\bibinfo{volume}{13920}. \bibinfo{publisher}{Springer},
  \bibinfo{pages}{313--326}.
\newblock
\urldef\tempurl%
\url{https://doi.org/10.1007/978-3-031-34960-7\_22}
\showDOI{\tempurl}


\bibitem[\protect\citeauthoryear{Cai, Jin, Zhao, and Chen}{Cai
  et~al\mbox{.}}{2023}]%
        {Cai2023Listen}
\bibfield{author}{\bibinfo{person}{Wanling Cai}, \bibinfo{person}{Yucheng Jin},
  \bibinfo{person}{Xianglin Zhao}, {and} \bibinfo{person}{Li Chen}.}
  \bibinfo{year}{2023}\natexlab{}.
\newblock \showarticletitle{{\textquotedblleft}Listen to music, listen to
  yourself{\textquotedblright}: Design of a conversational agent to support
  self-awareness while listening to music}. In
  \bibinfo{booktitle}{\emph{Proceedings of the 2023 {CHI} Conference on Human
  Factors in Computing Systems}}. \bibinfo{publisher}{{ACM}},
  \bibinfo{address}{New York, NY}, \bibinfo{pages}{119:1--119:19}.
\newblock
\urldef\tempurl%
\url{https://doi.org/10.1145/3544548.3581427}
\showDOI{\tempurl}


\bibitem[\protect\citeauthoryear{Casas, Mugellini, and Khaled}{Casas
  et~al\mbox{.}}{2018}]%
        {Casas2018Food}
\bibfield{author}{\bibinfo{person}{Jacky Casas}, \bibinfo{person}{Elena
  Mugellini}, {and} \bibinfo{person}{Omar~Abou Khaled}.}
  \bibinfo{year}{2018}\natexlab{}.
\newblock \showarticletitle{Food diary coaching chatbot}. In
  \bibinfo{booktitle}{\emph{Proceedings of the 2018 {ACM} International Joint
  Conference on Pervasive and Ubiquitous Computing and Wearable Computers}}.
  \bibinfo{publisher}{{ACM}}, \bibinfo{address}{New York, NY},
  \bibinfo{pages}{1676--1680}.
\newblock
\urldef\tempurl%
\url{https://doi.org/10.1145/3267305.3274191}
\showDOI{\tempurl}


\bibitem[\protect\citeauthoryear{Correia, dos Santos, and Passmore}{Correia
  et~al\mbox{.}}{2016}]%
        {correia2016understanding}
\bibfield{author}{\bibinfo{person}{Mara~Castro Correia},
  \bibinfo{person}{Nuno~Rebelo dos Santos}, {and} \bibinfo{person}{Jonathan
  Passmore}.} \bibinfo{year}{2016}\natexlab{}.
\newblock \showarticletitle{Understanding the Coach-Coachee-Client
  relationship: A conceptual framework for executive coaching}.
\newblock \bibinfo{journal}{\emph{International Coaching Psychology Review}}
  \bibinfo{volume}{11}, \bibinfo{number}{1} (\bibinfo{year}{2016}),
  \bibinfo{pages}{6--23}.
\newblock


\bibitem[\protect\citeauthoryear{Davis}{Davis}{1989}]%
        {Davis1989Perceived}
\bibfield{author}{\bibinfo{person}{Fred~D. Davis}.}
  \bibinfo{year}{1989}\natexlab{}.
\newblock \showarticletitle{Perceived usefulness, perceived ease of use, and
  user acceptance of information technology}.
\newblock \bibinfo{journal}{\emph{{MIS} Quarterly}} \bibinfo{volume}{13},
  \bibinfo{number}{3} (\bibinfo{year}{1989}), \bibinfo{pages}{319--340}.
\newblock
\urldef\tempurl%
\url{https://doi.org/10.2307/249008}
\showDOI{\tempurl}


\bibitem[\protect\citeauthoryear{Dwivedi et~al\mbox{.}}{Dwivedi
  et~al\mbox{.}}{2023}]%
        {Dwivedi2023Opinion}
\bibfield{author}{\bibinfo{person}{Yogesh~K. Dwivedi} {et~al\mbox{.}}}
  \bibinfo{year}{2023}\natexlab{}.
\newblock \showarticletitle{Opinion paper: {\textquotedblleft}So what if
  {ChatGPT} wrote it?{\textquotedblright} Multidisciplinary perspectives on
  opportunities, challenges and implications of generative conversational {AI}
  for research, practice and policy}.
\newblock \bibinfo{journal}{\emph{International Journal of Information
  Management}}  \bibinfo{volume}{71} (\bibinfo{year}{2023}),
  \bibinfo{pages}{102642}.
\newblock
\urldef\tempurl%
\url{https://doi.org/10.1016/j.ijinfomgt.2023.102642}
\showDOI{\tempurl}


\bibitem[\protect\citeauthoryear{Essel, Vlachopoulos, Tachie-Menson, Johnson,
  and Baah}{Essel et~al\mbox{.}}{2022}]%
        {Essel2022Impact}
\bibfield{author}{\bibinfo{person}{Harry~Barton Essel},
  \bibinfo{person}{Dimitrios Vlachopoulos}, \bibinfo{person}{Akosua
  Tachie-Menson}, \bibinfo{person}{Esi~Eduafua Johnson}, {and}
  \bibinfo{person}{Papa~Kwame Baah}.} \bibinfo{year}{2022}\natexlab{}.
\newblock \showarticletitle{The impact of a virtual teaching assistant
  (chatbot) on students{\textquotesingle} learning in Ghanaian higher
  education}.
\newblock \bibinfo{journal}{\emph{International Journal of Educational
  Technology in Higher Education}} \bibinfo{volume}{19}, \bibinfo{number}{1}
  (\bibinfo{year}{2022}).
\newblock
\urldef\tempurl%
\url{https://doi.org/10.1186/s41239-022-00362-6}
\showDOI{\tempurl}


\bibitem[\protect\citeauthoryear{Feine, Gnewuch, Morana, and Maedche}{Feine
  et~al\mbox{.}}{2019}]%
        {DBLP:conf/conversations/FeineGMM19}
\bibfield{author}{\bibinfo{person}{Jasper Feine}, \bibinfo{person}{Ulrich
  Gnewuch}, \bibinfo{person}{Stefan Morana}, {and} \bibinfo{person}{Alexander
  Maedche}.} \bibinfo{year}{2019}\natexlab{}.
\newblock \showarticletitle{Gender Bias in Chatbot Design}. In
  \bibinfo{booktitle}{\emph{Proceedings of the 3rd International Workshop on
  Chatbot Research and Design}}. \bibinfo{publisher}{Springer},
  \bibinfo{pages}{79--93}.
\newblock
\urldef\tempurl%
\url{https://doi.org/10.1007/978-3-030-39540-7\_6}
\showDOI{\tempurl}


\bibitem[\protect\citeauthoryear{Feldman and Lankau}{Feldman and
  Lankau}{2005}]%
        {doi:10.1177/0149206305279599}
\bibfield{author}{\bibinfo{person}{Daniel~C. Feldman} {and}
  \bibinfo{person}{Melenie~J. Lankau}.} \bibinfo{year}{2005}\natexlab{}.
\newblock \showarticletitle{Executive coaching: A review and agenda for future
  research}.
\newblock \bibinfo{journal}{\emph{Journal of Management}} \bibinfo{volume}{31},
  \bibinfo{number}{6} (\bibinfo{year}{2005}), \bibinfo{pages}{829--848}.
\newblock
\urldef\tempurl%
\url{https://doi.org/10.1177/0149206305279599}
\showDOI{\tempurl}


\bibitem[\protect\citeauthoryear{F{\o}lstad, Skjuve, and Brandtzaeg}{F{\o}lstad
  et~al\mbox{.}}{2019}]%
        {Flstad2019Different}
\bibfield{author}{\bibinfo{person}{Asbj{\o}rn F{\o}lstad},
  \bibinfo{person}{Marita Skjuve}, {and} \bibinfo{person}{Petter~Bae
  Brandtzaeg}.} \bibinfo{year}{2019}\natexlab{}.
\newblock \showarticletitle{Different chatbots for different purposes: Towards
  a typology of chatbots to understand interaction design}.
\newblock In \bibinfo{booktitle}{\emph{Proceedings of the 2019 International
  Conference on Internet Science}}. \bibinfo{publisher}{Springer International
  Publishing}, \bibinfo{address}{Cham, Switzerland}, \bibinfo{pages}{145--156}.
\newblock
\urldef\tempurl%
\url{https://doi.org/10.1007/978-3-030-17705-8_13}
\showDOI{\tempurl}


\bibitem[\protect\citeauthoryear{Ford and Bryan{-}Kinns}{Ford and
  Bryan{-}Kinns}{2023}]%
        {DBLP:conf/chi/0002B23}
\bibfield{author}{\bibinfo{person}{Corey Ford} {and} \bibinfo{person}{Nick
  Bryan{-}Kinns}.} \bibinfo{year}{2023}\natexlab{}.
\newblock \showarticletitle{Towards a reflection in creative experience
  questionnaire}. In \bibinfo{booktitle}{\emph{Proceedings of the 2023 {ACM}
  {CHI} Conference on Human Factors in Computing Systems}}.
  \bibinfo{publisher}{{ACM}}, \bibinfo{address}{New York, NY},
  \bibinfo{pages}{763:1--763:16}.
\newblock
\urldef\tempurl%
\url{https://doi.org/10.1145/3544548.3581077}
\showDOI{\tempurl}


\bibitem[\protect\citeauthoryear{Gabrielli, Rizzi, Carbone, and
  Donisi}{Gabrielli et~al\mbox{.}}{2020}]%
        {Gabrielli2020}
\bibfield{author}{\bibinfo{person}{Silvia Gabrielli}, \bibinfo{person}{Silvia
  Rizzi}, \bibinfo{person}{Sara Carbone}, {and} \bibinfo{person}{Valeria
  Donisi}.} \bibinfo{year}{2020}\natexlab{}.
\newblock \showarticletitle{A chatbot-based coaching intervention for
  adolescents to promote life skills: Pilot study}.
\newblock \bibinfo{journal}{\emph{{JMIR} Human Factors}} \bibinfo{volume}{7},
  \bibinfo{number}{1} (\bibinfo{year}{2020}), \bibinfo{pages}{e16762}.
\newblock
\urldef\tempurl%
\url{https://doi.org/10.2196/16762}
\showDOI{\tempurl}


\bibitem[\protect\citeauthoryear{Grant}{Grant}{2012}]%
        {Grant2012Making}
\bibfield{author}{\bibinfo{person}{Anthony~M. Grant}.}
  \bibinfo{year}{2012}\natexlab{}.
\newblock \showarticletitle{Making positive change: A randomized study
  comparing solution-focused vs. problem-focused coaching questions}.
\newblock \bibinfo{journal}{\emph{Journal of Systemic Therapies}}
  \bibinfo{volume}{31}, \bibinfo{number}{2} (\bibinfo{year}{2012}),
  \bibinfo{pages}{21--35}.
\newblock
\urldef\tempurl%
\url{https://doi.org/10.1521/jsyt.2012.31.2.21}
\showDOI{\tempurl}


\bibitem[\protect\citeauthoryear{Gra{\ss}mann and Schermuly}{Gra{\ss}mann and
  Schermuly}{2021}]%
        {grassmann2021coaching}
\bibfield{author}{\bibinfo{person}{Carolin Gra{\ss}mann} {and}
  \bibinfo{person}{Carsten~C Schermuly}.} \bibinfo{year}{2021}\natexlab{}.
\newblock \showarticletitle{Coaching with artificial intelligence: Concepts and
  capabilities}.
\newblock \bibinfo{journal}{\emph{Human Resource Development Review}}
  \bibinfo{volume}{20}, \bibinfo{number}{1} (\bibinfo{year}{2021}),
  \bibinfo{pages}{106--126}.
\newblock


\bibitem[\protect\citeauthoryear{Greenfield and Hengen}{Greenfield and
  Hengen}{2004}]%
        {Greenfield2004}
\bibfield{author}{\bibinfo{person}{Daniel~P. Greenfield} {and}
  \bibinfo{person}{William~K. Hengen}.} \bibinfo{year}{2004}\natexlab{}.
\newblock \showarticletitle{Confidentiality in Coaching}.
\newblock \bibinfo{journal}{\emph{Consulting to Management}}
  \bibinfo{volume}{15}, \bibinfo{number}{1}, \bibinfo{pages}{9--14}.
\newblock


\bibitem[\protect\citeauthoryear{Hawkins}{Hawkins}{2008}]%
        {Hawkins2008}
\bibfield{author}{\bibinfo{person}{Peter Hawkins}.}
  \bibinfo{year}{2008}\natexlab{}.
\newblock \showarticletitle{The coaching profession: some of the key
  challenges}.
\newblock \bibinfo{journal}{\emph{Coaching: An International Journal of Theory,
  Research and Practice}} \bibinfo{volume}{1}, \bibinfo{number}{1}
  (\bibinfo{year}{2008}), \bibinfo{pages}{28--38}.
\newblock
\urldef\tempurl%
\url{https://doi.org/10.1080/17521880701878174}
\showDOI{\tempurl}


\bibitem[\protect\citeauthoryear{Joo}{Joo}{2005}]%
        {Joo2005Executive}
\bibfield{author}{\bibinfo{person}{Baek-Kyoo~(Brian) Joo}.}
  \bibinfo{year}{2005}\natexlab{}.
\newblock \showarticletitle{Executive Coaching: A Conceptual Framework From an
  Integrative Review of Practice and Research}.
\newblock \bibinfo{journal}{\emph{Human Resource Development Review}}
  \bibinfo{volume}{4}, \bibinfo{number}{4} (\bibinfo{year}{2005}),
  \bibinfo{pages}{462--488}.
\newblock
\urldef\tempurl%
\url{https://doi.org/10.1177/1534484305280866}
\showDOI{\tempurl}


\bibitem[\protect\citeauthoryear{J{\"{o}}rke, Sefidgar, Massachi, Suh, and
  Ramos}{J{\"{o}}rke et~al\mbox{.}}{2023}]%
        {DBLP:conf/iui/JorkeSMSR23}
\bibfield{author}{\bibinfo{person}{Matthew J{\"{o}}rke},
  \bibinfo{person}{Yasaman~S. Sefidgar}, \bibinfo{person}{Talie Massachi},
  \bibinfo{person}{Jina Suh}, {and} \bibinfo{person}{Gonzalo~A. Ramos}.}
  \bibinfo{year}{2023}\natexlab{}.
\newblock \showarticletitle{Pearl: {A} technology probe for machine-assisted
  reflection on personal data}. In \bibinfo{booktitle}{\emph{Proceedings of the
  28th {ACM} International Conference on Intelligent User Interfaces}}.
  \bibinfo{publisher}{{ACM}}, \bibinfo{address}{New York, NY},
  \bibinfo{pages}{902--918}.
\newblock
\urldef\tempurl%
\url{https://doi.org/10.1145/3581641.3584054}
\showDOI{\tempurl}


\bibitem[\protect\citeauthoryear{Kamali, Angelini, Caon, Andreoni, Khaled, and
  Mugellini}{Kamali et~al\mbox{.}}{2018}]%
        {DBLP:conf/huc/KamaliACAKM18}
\bibfield{author}{\bibinfo{person}{Mira~El Kamali}, \bibinfo{person}{Leonardo
  Angelini}, \bibinfo{person}{Maurizio Caon}, \bibinfo{person}{Giuseppe
  Andreoni}, \bibinfo{person}{Omar~Abou Khaled}, {and} \bibinfo{person}{Elena
  Mugellini}.} \bibinfo{year}{2018}\natexlab{}.
\newblock \showarticletitle{Towards the {NESTORE} e-Coach: A tangible and
  embodied conversational agent for older adults}. In
  \bibinfo{booktitle}{\emph{Proceedings of the 2018 {ACM} International Joint
  Conference on Pervasive and Ubiquitous Computing and Wearable Computers}}.
  \bibinfo{publisher}{{ACM}}, \bibinfo{address}{New York, NY},
  \bibinfo{pages}{1656--1663}.
\newblock
\urldef\tempurl%
\url{https://doi.org/10.1145/3267305.3274188}
\showDOI{\tempurl}


\bibitem[\protect\citeauthoryear{Kilburg}{Kilburg}{1997}]%
        {Kilburg1997}
\bibfield{author}{\bibinfo{person}{Richard~R. Kilburg}.}
  \bibinfo{year}{1997}\natexlab{}.
\newblock \showarticletitle{Coaching and executive character: Core problems and
  basic approaches}.
\newblock \bibinfo{journal}{\emph{Consulting Psychology Journal: Practice and
  Research}} \bibinfo{volume}{49}, \bibinfo{number}{4} (\bibinfo{year}{1997}),
  \bibinfo{pages}{281--299}.
\newblock
\urldef\tempurl%
\url{https://doi.org/10.1037/1061-4087.49.4.281}
\showDOI{\tempurl}


\bibitem[\protect\citeauthoryear{Kim, Lee, Kim, Jo, Yoo, Hwang, Kang, and
  Song}{Kim et~al\mbox{.}}{2020}]%
        {DBLP:journals/imwut/KimLKJYHKS20}
\bibfield{author}{\bibinfo{person}{Wonjung Kim}, \bibinfo{person}{Seungchul
  Lee}, \bibinfo{person}{Seonghoon Kim}, \bibinfo{person}{Sungbin Jo},
  \bibinfo{person}{Chungkuk Yoo}, \bibinfo{person}{Inseok Hwang},
  \bibinfo{person}{Seungwoo Kang}, {and} \bibinfo{person}{Junehwa Song}.}
  \bibinfo{year}{2020}\natexlab{}.
\newblock \showarticletitle{Dyadic Mirror: Everyday Second-person Live-view for
  Empathetic Reflection upon Parent-child Interaction}.
\newblock \bibinfo{journal}{\emph{Proc. {ACM} Interact. Mob. Wearable
  Ubiquitous Technol.}} \bibinfo{volume}{4}, \bibinfo{number}{3}
  (\bibinfo{year}{2020}), \bibinfo{pages}{86:1--86:29}.
\newblock
\urldef\tempurl%
\url{https://doi.org/10.1145/3411815}
\showDOI{\tempurl}


\bibitem[\protect\citeauthoryear{Kocielnik, Xiao, Avrahami, and
  Hsieh}{Kocielnik et~al\mbox{.}}{2018}]%
        {DBLP:journals/imwut/KocielnikXAH18}
\bibfield{author}{\bibinfo{person}{Rafal Kocielnik}, \bibinfo{person}{Lillian
  Xiao}, \bibinfo{person}{Daniel Avrahami}, {and} \bibinfo{person}{Gary
  Hsieh}.} \bibinfo{year}{2018}\natexlab{}.
\newblock \showarticletitle{Reflection Companion: {A} Conversational System for
  Engaging Users in Reflection on Physical Activity}.
\newblock \bibinfo{journal}{\emph{Proc. {ACM} Interact. Mob. Wearable
  Ubiquitous Technol.}} \bibinfo{volume}{2}, \bibinfo{number}{2}
  (\bibinfo{year}{2018}), \bibinfo{pages}{70:1--70:26}.
\newblock
\urldef\tempurl%
\url{https://doi.org/10.1145/3214273}
\showDOI{\tempurl}


\bibitem[\protect\citeauthoryear{Kojima, Gu, Reid, Matsuo, and Iwasawa}{Kojima
  et~al\mbox{.}}{2022}]%
        {DBLP:conf/nips/KojimaGRMI22}
\bibfield{author}{\bibinfo{person}{Takeshi Kojima},
  \bibinfo{person}{Shixiang~Shane Gu}, \bibinfo{person}{Machel Reid},
  \bibinfo{person}{Yutaka Matsuo}, {and} \bibinfo{person}{Yusuke Iwasawa}.}
  \bibinfo{year}{2022}\natexlab{}.
\newblock \showarticletitle{Large Language Models are Zero-Shot Reasoners}. In
  \bibinfo{booktitle}{\emph{Proceedings of the 36th Annual Conference on Neural
  Information Processing Systems}}. \bibinfo{publisher}{Curran Associates,
  Inc.}, \bibinfo{address}{Red Hook, NY}, \bibinfo{pages}{22199--22213}.
\newblock


\bibitem[\protect\citeauthoryear{Kuhail, Alturki, Alramlawi, and
  Alhejori}{Kuhail et~al\mbox{.}}{2023}]%
        {DBLP:journals/eait/KuhailAAA23}
\bibfield{author}{\bibinfo{person}{Mohammad~Amin Kuhail},
  \bibinfo{person}{Nazik Alturki}, \bibinfo{person}{Salwa Alramlawi}, {and}
  \bibinfo{person}{Kholood Alhejori}.} \bibinfo{year}{2023}\natexlab{}.
\newblock \showarticletitle{Interacting with educational chatbots: {A}
  systematic review}.
\newblock \bibinfo{journal}{\emph{Education and Information Technologies}}
  \bibinfo{volume}{28}, \bibinfo{number}{1} (\bibinfo{year}{2023}),
  \bibinfo{pages}{973--1018}.
\newblock
\urldef\tempurl%
\url{https://doi.org/10.1007/s10639-022-11177-3}
\showDOI{\tempurl}


\bibitem[\protect\citeauthoryear{Lee, Ackermans, van As, Chang, Lucas, and
  IJsselsteijn}{Lee et~al\mbox{.}}{2019}]%
        {DBLP:conf/chi/LeeAACLI19}
\bibfield{author}{\bibinfo{person}{Minha Lee}, \bibinfo{person}{Sander
  Ackermans}, \bibinfo{person}{Nena van As}, \bibinfo{person}{Hanwen Chang},
  \bibinfo{person}{Enzo Lucas}, {and} \bibinfo{person}{Wijnand~A.
  IJsselsteijn}.} \bibinfo{year}{2019}\natexlab{}.
\newblock \showarticletitle{Caring for Vincent: {A} Chatbot for
  Self-Compassion}. In \bibinfo{booktitle}{\emph{Proceedings of the 2019 {ACM}
  {CHI} Conference on Human Factors in Computing Systems}}.
  \bibinfo{publisher}{{ACM}}, \bibinfo{pages}{702:1--702:13}.
\newblock
\urldef\tempurl%
\url{https://doi.org/10.1145/3290605.3300932}
\showDOI{\tempurl}


\bibitem[\protect\citeauthoryear{Li, Liang, Le, LC, and Luo}{Li
  et~al\mbox{.}}{2023}]%
        {DBLP:conf/cui/LiLLLL23}
\bibfield{author}{\bibinfo{person}{Zhuoyang Li}, \bibinfo{person}{Minhui
  Liang}, \bibinfo{person}{Hai~Trung Le}, \bibinfo{person}{Ray LC}, {and}
  \bibinfo{person}{Yuhan Luo}.} \bibinfo{year}{2023}\natexlab{}.
\newblock \showarticletitle{Exploring Design Opportunities for Reflective
  Conversational Agents to Reduce Compulsive Smartphone Use}. In
  \bibinfo{booktitle}{\emph{Proceedings of the 5th International Conference on
  Conversational User Interfaces}}. \bibinfo{publisher}{{ACM}},
  \bibinfo{pages}{37:1--37:6}.
\newblock
\urldef\tempurl%
\url{https://doi.org/10.1145/3571884.3604305}
\showDOI{\tempurl}


\bibitem[\protect\citeauthoryear{Liu, Hou, Tu, Wang, and Hwang}{Liu
  et~al\mbox{.}}{2021}]%
        {Liu2021}
\bibfield{author}{\bibinfo{person}{Chenchen Liu}, \bibinfo{person}{Jierui Hou},
  \bibinfo{person}{Yun-Fang Tu}, \bibinfo{person}{Youmei Wang}, {and}
  \bibinfo{person}{Gwo-Jen Hwang}.} \bibinfo{year}{2021}\natexlab{}.
\newblock \showarticletitle{Incorporating a Reflective Thinking Promoting
  Mechanism into Artificial Intelligence-Supported English Writing
  Environments}.
\newblock \bibinfo{journal}{\emph{Interactive Learning Environments}}
  \bibinfo{volume}{31}, \bibinfo{number}{9} (\bibinfo{year}{2021}),
  \bibinfo{pages}{5614--5632}.
\newblock
\urldef\tempurl%
\url{https://doi.org/10.1080/10494820.2021.2012812}
\showDOI{\tempurl}


\bibitem[\protect\citeauthoryear{MacKie}{MacKie}{2015}]%
        {MacKie2015Effects}
\bibfield{author}{\bibinfo{person}{Doug MacKie}.}
  \bibinfo{year}{2015}\natexlab{}.
\newblock \showarticletitle{The effects of coachee readiness and core
  self-evaluations on leadership coaching outcomes: A controlled trial}.
\newblock \bibinfo{journal}{\emph{Coaching: An International Journal of Theory,
  Research and Practice}} \bibinfo{volume}{8}, \bibinfo{number}{2}
  (\bibinfo{year}{2015}), \bibinfo{pages}{120--136}.
\newblock
\urldef\tempurl%
\url{https://doi.org/10.1080/17521882.2015.1019532}
\showDOI{\tempurl}


\bibitem[\protect\citeauthoryear{Mai, Wolff, Richert, and Preusser}{Mai
  et~al\mbox{.}}{2021}]%
        {DBLP:conf/hci/MaiWRP21}
\bibfield{author}{\bibinfo{person}{Vanessa Mai}, \bibinfo{person}{Annika
  Wolff}, \bibinfo{person}{Anja Richert}, {and} \bibinfo{person}{Ivonne
  Preusser}.} \bibinfo{year}{2021}\natexlab{}.
\newblock \showarticletitle{Accompanying reflection processes by an AI-based
  StudiCoachBot: {A} study on rapport building in human-machine coaching using
  self disclosure}. In \bibinfo{booktitle}{\emph{Proceedings of the 23rd {HCI}
  International Conference - Late Breaking Papers}}.
  \bibinfo{publisher}{Springer}, \bibinfo{address}{Cham, Switzerland},
  \bibinfo{pages}{439--457}.
\newblock
\urldef\tempurl%
\url{https://doi.org/10.1007/978-3-030-90328-2\_29}
\showDOI{\tempurl}


\bibitem[\protect\citeauthoryear{Mitchell, Elhadad, and Mamykina}{Mitchell
  et~al\mbox{.}}{2022}]%
        {DBLP:conf/chi/MitchellEM22}
\bibfield{author}{\bibinfo{person}{Elliot~G. Mitchell}, \bibinfo{person}{Noemie
  Elhadad}, {and} \bibinfo{person}{Lena Mamykina}.}
  \bibinfo{year}{2022}\natexlab{}.
\newblock \showarticletitle{Examining {AI} methods for micro-coaching dialogs}.
  In \bibinfo{booktitle}{\emph{Proceedings of the 2022 {ACM} {CHI} Conference
  on Human Factors in Computing Systems}}. \bibinfo{publisher}{{ACM}},
  \bibinfo{address}{New York, NY}, \bibinfo{pages}{440:1--440:24}.
\newblock
\urldef\tempurl%
\url{https://doi.org/10.1145/3491102.3501886}
\showDOI{\tempurl}


\bibitem[\protect\citeauthoryear{Mitchell, Maimone, Cassells, Tobin, Davidson,
  Smaldone, and Mamykina}{Mitchell et~al\mbox{.}}{2021}]%
        {DBLP:journals/pacmhci/MitchellMCTDSM21}
\bibfield{author}{\bibinfo{person}{Elliot~G. Mitchell}, \bibinfo{person}{Rosa
  Maimone}, \bibinfo{person}{Andrea Cassells}, \bibinfo{person}{Jonathan~N.
  Tobin}, \bibinfo{person}{Patricia~G. Davidson}, \bibinfo{person}{Arlene~M.
  Smaldone}, {and} \bibinfo{person}{Lena Mamykina}.}
  \bibinfo{year}{2021}\natexlab{}.
\newblock \showarticletitle{Automated vs. human health coaching: Exploring
  participant and practitioner experiences}.
\newblock \bibinfo{journal}{\emph{Proceedings of the ACM on Human-Computer
  Interaction}} \bibinfo{volume}{5}, \bibinfo{number}{{CSCW1}}
  (\bibinfo{year}{2021}), \bibinfo{pages}{99:1--99:37}.
\newblock
\urldef\tempurl%
\url{https://doi.org/10.1145/3449173}
\showDOI{\tempurl}


\bibitem[\protect\citeauthoryear{Moen and Skaalvik}{Moen and Skaalvik}{2009}]%
        {moen2009effect}
\bibfield{author}{\bibinfo{person}{Frode Moen} {and} \bibinfo{person}{Einar
  Skaalvik}.} \bibinfo{year}{2009}\natexlab{}.
\newblock \showarticletitle{The effect from executive coaching on performance
  psychology}.
\newblock \bibinfo{journal}{\emph{International Journal of Evidence Based
  Coaching \& Mentoring}} \bibinfo{volume}{7}, \bibinfo{number}{2}
  (\bibinfo{year}{2009}), \bibinfo{pages}{31--49}.
\newblock


\bibitem[\protect\citeauthoryear{Mols, van~den Hoven, and Eggen}{Mols
  et~al\mbox{.}}{2016}]%
        {Mols2016Informing}
\bibfield{author}{\bibinfo{person}{Ine Mols}, \bibinfo{person}{Elise van~den
  Hoven}, {and} \bibinfo{person}{Berry Eggen}.}
  \bibinfo{year}{2016}\natexlab{}.
\newblock \showarticletitle{Informing design for reflection}. In
  \bibinfo{booktitle}{\emph{Proceedings of the 9th Nordic Conference on
  Human-Computer Interaction}}. \bibinfo{publisher}{{ACM}},
  \bibinfo{address}{New York, NY}, \bibinfo{pages}{1--10}.
\newblock
\urldef\tempurl%
\url{https://doi.org/10.1145/2971485.2971494}
\showDOI{\tempurl}


\bibitem[\protect\citeauthoryear{Moussa-Inaty}{Moussa-Inaty}{2015}]%
        {Moussa-Inaty2015}
\bibfield{author}{\bibinfo{person}{Jase Moussa-Inaty}.}
  \bibinfo{year}{2015}\natexlab{}.
\newblock \showarticletitle{Reflective Writing through the Use of Guiding
  Questions}.
\newblock \bibinfo{journal}{\emph{International Journal of Teaching and
  Learning in Higher Education}} \bibinfo{volume}{27}, \bibinfo{number}{1}
  (\bibinfo{year}{2015}), \bibinfo{pages}{104--113}.
\newblock


\bibitem[\protect\citeauthoryear{Narain, Quach, Davey, Park, Breazeal, and
  Picard}{Narain et~al\mbox{.}}{2020}]%
        {Narain2020Promoting}
\bibfield{author}{\bibinfo{person}{Jaya Narain}, \bibinfo{person}{Tina Quach},
  \bibinfo{person}{Monique Davey}, \bibinfo{person}{Hae~Won Park},
  \bibinfo{person}{Cynthia Breazeal}, {and} \bibinfo{person}{Rosalind Picard}.}
  \bibinfo{year}{2020}\natexlab{}.
\newblock \showarticletitle{Promoting wellbeing with sunny, a chatbot that
  facilitates positive messages within social groups}. In
  \bibinfo{booktitle}{\emph{Extended Abstracts of the 2020 {ACM} {CHI}
  Conference on Human Factors in Computing Systems}}.
  \bibinfo{publisher}{{ACM}}, \bibinfo{address}{New York, NY},
  \bibinfo{pages}{1--8}.
\newblock
\urldef\tempurl%
\url{https://doi.org/10.1145/3334480.3383062}
\showDOI{\tempurl}


\bibitem[\protect\citeauthoryear{Ouyang, Wu, Jiang, Almeida, Wainwright,
  Mishkin, Zhang, Agarwal, Slama, Ray, Schulman, Hilton, Kelton, Miller,
  Simens, Askell, Welinder, Christiano, Leike, and Lowe}{Ouyang
  et~al\mbox{.}}{2022}]%
        {DBLP:conf/nips/Ouyang0JAWMZASR22}
\bibfield{author}{\bibinfo{person}{Long Ouyang}, \bibinfo{person}{Jeffrey Wu},
  \bibinfo{person}{Xu Jiang}, \bibinfo{person}{Diogo Almeida},
  \bibinfo{person}{Carroll~L. Wainwright}, \bibinfo{person}{Pamela Mishkin},
  \bibinfo{person}{Chong Zhang}, \bibinfo{person}{Sandhini Agarwal},
  \bibinfo{person}{Katarina Slama}, \bibinfo{person}{Alex Ray},
  \bibinfo{person}{John Schulman}, \bibinfo{person}{Jacob Hilton},
  \bibinfo{person}{Fraser Kelton}, \bibinfo{person}{Luke Miller},
  \bibinfo{person}{Maddie Simens}, \bibinfo{person}{Amanda Askell},
  \bibinfo{person}{Peter Welinder}, \bibinfo{person}{Paul~F. Christiano},
  \bibinfo{person}{Jan Leike}, {and} \bibinfo{person}{Ryan Lowe}.}
  \bibinfo{year}{2022}\natexlab{}.
\newblock \showarticletitle{Training language models to follow instructions
  with human feedback}. In \bibinfo{booktitle}{\emph{Proceedings of the 36th
  Annual Conference on Neural Information Processing Systems}}.
  \bibinfo{publisher}{Curran Associates, Inc.}, \bibinfo{address}{Red Hook,
  NY}, \bibinfo{pages}{27730--27744}.
\newblock


\bibitem[\protect\citeauthoryear{Passmore and Tee}{Passmore and Tee}{2023}]%
        {Passmore2023}
\bibfield{author}{\bibinfo{person}{Jonathan Passmore} {and}
  \bibinfo{person}{David Tee}.} \bibinfo{year}{2023}\natexlab{}.
\newblock \showarticletitle{Can chatbots replace human coaches? Issues and
  dilemmas for the coaching profession, coaching clients and for
  organisations}.
\newblock \bibinfo{journal}{\emph{The Coaching Psychologist}}
  \bibinfo{volume}{19}, \bibinfo{number}{1} (\bibinfo{year}{2023}),
  \bibinfo{pages}{47--54}.
\newblock
\urldef\tempurl%
\url{https://doi.org/10.53841/bpstcp.2023.19.1.47}
\showDOI{\tempurl}


\bibitem[\protect\citeauthoryear{Qiu, Yuan, Bi, Huang, and You}{Qiu
  et~al\mbox{.}}{2023}]%
        {DBLP:conf/chi/QiuYBHY23}
\bibfield{author}{\bibinfo{person}{Xiang{-}Zhi Qiu},
  \bibinfo{person}{Tina~Chien{-}Wen Yuan}, \bibinfo{person}{Nanyi Bi},
  \bibinfo{person}{Ming{-}Chyi Huang}, {and} \bibinfo{person}{Chuang{-}Wen
  You}.} \bibinfo{year}{2023}\natexlab{}.
\newblock \showarticletitle{Exploring the challenges and opportunities in
  developing systems to improve alcohol use disorder through chatbot
  technology}. In \bibinfo{booktitle}{\emph{Extended Abstracts of the 2023
  {ACM} {CHI} Conference on Human Factors in Computing Systems}}.
  \bibinfo{publisher}{{ACM}}, \bibinfo{address}{New York, NY},
  \bibinfo{pages}{123:1--123:5}.
\newblock
\urldef\tempurl%
\url{https://doi.org/10.1145/3544549.3585635}
\showDOI{\tempurl}


\bibitem[\protect\citeauthoryear{Rutjes, Willemsen, and IJsselsteijn}{Rutjes
  et~al\mbox{.}}{2019}]%
        {DBLP:conf/chi/RutjesWI19}
\bibfield{author}{\bibinfo{person}{Heleen Rutjes}, \bibinfo{person}{Martijn~C.
  Willemsen}, {and} \bibinfo{person}{Wijnand~A. IJsselsteijn}.}
  \bibinfo{year}{2019}\natexlab{}.
\newblock \showarticletitle{Beyond behavior: The coach's perspective on
  technology in health coaching}. In \bibinfo{booktitle}{\emph{Proceedings of
  the 2019 {CHI} Conference on Human Factors in Computing Systems}}.
  \bibinfo{publisher}{{ACM}}, \bibinfo{address}{New York, NY},
  \bibinfo{pages}{670}.
\newblock
\urldef\tempurl%
\url{https://doi.org/10.1145/3290605.3300900}
\showDOI{\tempurl}


\bibitem[\protect\citeauthoryear{Ryan, Dockray, and Linehan}{Ryan
  et~al\mbox{.}}{2022}]%
        {DBLP:conf/chi/RyanDL22}
\bibfield{author}{\bibinfo{person}{Kathleen Ryan}, \bibinfo{person}{Samantha
  Dockray}, {and} \bibinfo{person}{Conor Linehan}.}
  \bibinfo{year}{2022}\natexlab{}.
\newblock \showarticletitle{Understanding how {eHealth} coaches tailor support
  for weight loss: Towards the design of person-centered coaching systems}. In
  \bibinfo{booktitle}{\emph{Proceedings of the 2022 {ACM} {CHI} Conference on
  Human Factors in Computing Systems}}. \bibinfo{publisher}{{ACM}},
  \bibinfo{address}{New York, NY}, \bibinfo{pages}{285:1--285:16}.
\newblock
\urldef\tempurl%
\url{https://doi.org/10.1145/3491102.3501864}
\showDOI{\tempurl}


\bibitem[\protect\citeauthoryear{Sadek, Calvo, and Mougenot}{Sadek
  et~al\mbox{.}}{2023}]%
        {DBLP:conf/cui/SadekCM23}
\bibfield{author}{\bibinfo{person}{Malak Sadek}, \bibinfo{person}{Rafael~A.
  Calvo}, {and} \bibinfo{person}{C{\'{e}}line Mougenot}.}
  \bibinfo{year}{2023}\natexlab{}.
\newblock \showarticletitle{Trends, Challenges and Processes in Conversational
  Agent Design: Exploring Practitioners' Views through Semi-Structured
  Interviews}. In \bibinfo{booktitle}{\emph{Proceedings of the 5th
  International Conference on Conversational User Interfaces}}.
  \bibinfo{publisher}{{ACM}}, \bibinfo{pages}{13:1--13:10}.
\newblock
\urldef\tempurl%
\url{https://doi.org/10.1145/3571884.3597143}
\showDOI{\tempurl}


\bibitem[\protect\citeauthoryear{Samrose, McDuff, Sim, Suh, Rowan, Hernandez,
  Rintel, Moynihan, and Czerwinski}{Samrose et~al\mbox{.}}{2021}]%
        {DBLP:conf/chi/SamroseMSSRHRMC21}
\bibfield{author}{\bibinfo{person}{Samiha Samrose}, \bibinfo{person}{Daniel
  McDuff}, \bibinfo{person}{Robert Sim}, \bibinfo{person}{Jina Suh},
  \bibinfo{person}{Kael Rowan}, \bibinfo{person}{Javier Hernandez},
  \bibinfo{person}{Sean Rintel}, \bibinfo{person}{Kevin Moynihan}, {and}
  \bibinfo{person}{Mary Czerwinski}.} \bibinfo{year}{2021}\natexlab{}.
\newblock \showarticletitle{MeetingCoach: An Intelligent Dashboard for
  Supporting Effective {\&} Inclusive Meetings}. In
  \bibinfo{booktitle}{\emph{Proceedings of the 2021 {ACM} {CHI} Conference on
  Human Factors in Computing Systems}}. \bibinfo{publisher}{{ACM}},
  \bibinfo{pages}{252:1--252:13}.
\newblock
\urldef\tempurl%
\url{https://doi.org/10.1145/3411764.3445615}
\showDOI{\tempurl}


\bibitem[\protect\citeauthoryear{Strauss and Corbin}{Strauss and
  Corbin}{1990}]%
        {StrCor90}
\bibfield{author}{\bibinfo{person}{Anselm~L Strauss} {and}
  \bibinfo{person}{Juliet~M Corbin}.} \bibinfo{year}{1990}\natexlab{}.
\newblock \bibinfo{booktitle}{\emph{Basics of qualitative research: Grounded
  theory procedures and techniques}}.
\newblock \bibinfo{publisher}{Sage Publications}, \bibinfo{address}{Newbury
  Park, CA}.
\newblock


\bibitem[\protect\citeauthoryear{Susing, Green, and Grant}{Susing
  et~al\mbox{.}}{2011}]%
        {susing2011potential}
\bibfield{author}{\bibinfo{person}{Ingo Susing}, \bibinfo{person}{Suzy Green},
  {and} \bibinfo{person}{Anthony~M Grant}.} \bibinfo{year}{2011}\natexlab{}.
\newblock \showarticletitle{The potential use of the authenticity scale as an
  outcome measure in executive coaching}.
\newblock \bibinfo{journal}{\emph{The Coaching Psychologist}}
  \bibinfo{volume}{7}, \bibinfo{number}{1} (\bibinfo{year}{2011}),
  \bibinfo{pages}{16--25}.
\newblock


\bibitem[\protect\citeauthoryear{Terblanche and Cilliers}{Terblanche and
  Cilliers}{2020}]%
        {Terblanche2020Factors}
\bibfield{author}{\bibinfo{person}{Nicky Terblanche} {and}
  \bibinfo{person}{Danie Cilliers}.} \bibinfo{year}{2020}\natexlab{}.
\newblock \showarticletitle{Factors that influence users' adoption of being
  coached by an artificial intelligence coach}.
\newblock \bibinfo{journal}{\emph{Philosophy of Coaching: An International
  Journal}} \bibinfo{volume}{5}, \bibinfo{number}{1} (\bibinfo{year}{2020}),
  \bibinfo{pages}{61--70}.
\newblock
\urldef\tempurl%
\url{https://doi.org/10.22316/poc/05.1.06}
\showDOI{\tempurl}


\bibitem[\protect\citeauthoryear{Terblanche, Molyn, Williams, and
  Maritz}{Terblanche et~al\mbox{.}}{2022}]%
        {Terblanche2022Performance}
\bibfield{author}{\bibinfo{person}{Nicky Terblanche}, \bibinfo{person}{Joanna
  Molyn}, \bibinfo{person}{Kevin Williams}, {and} \bibinfo{person}{Jeanette
  Maritz}.} \bibinfo{year}{2022}\natexlab{}.
\newblock \showarticletitle{Performance matters: Students' perceptions of
  artificial intelligence coach adoption factors}.
\newblock \bibinfo{journal}{\emph{Coaching: An International Journal of Theory,
  Research and Practice}} \bibinfo{volume}{16}, \bibinfo{number}{1}
  (\bibinfo{year}{2022}), \bibinfo{pages}{100--114}.
\newblock
\urldef\tempurl%
\url{https://doi.org/10.1080/17521882.2022.2094278}
\showDOI{\tempurl}


\bibitem[\protect\citeauthoryear{Venkatesh, Morris, Davis, and Davis}{Venkatesh
  et~al\mbox{.}}{2003}]%
        {Venkatesh2003User}
\bibfield{author}{\bibinfo{person}{Viswanath Venkatesh},
  \bibinfo{person}{Michael~G. Morris}, \bibinfo{person}{Gordon~B. Davis}, {and}
  \bibinfo{person}{Fred~D. Davis}.} \bibinfo{year}{2003}\natexlab{}.
\newblock \showarticletitle{User acceptance of information technology: Toward a
  unified view}.
\newblock \bibinfo{journal}{\emph{{MIS} Quarterly}} \bibinfo{volume}{27},
  \bibinfo{number}{3} (\bibinfo{year}{2003}), \bibinfo{pages}{425--478}.
\newblock
\urldef\tempurl%
\url{https://doi.org/10.2307/30036540}
\showDOI{\tempurl}


\bibitem[\protect\citeauthoryear{Wagener, Reicherts, Zargham, Bartlomiejczyk,
  Scott, Wang, Bentvelzen, Stefanidi, Mildner, Rogers, and Niess}{Wagener
  et~al\mbox{.}}{2023}]%
        {DBLP:conf/chi/WagenerRZBSWBSM23}
\bibfield{author}{\bibinfo{person}{Nadine Wagener}, \bibinfo{person}{Leon
  Reicherts}, \bibinfo{person}{Nima Zargham}, \bibinfo{person}{Natalia
  Bartlomiejczyk}, \bibinfo{person}{Ava~Elizabeth Scott},
  \bibinfo{person}{Katherine Wang}, \bibinfo{person}{Marit Bentvelzen},
  \bibinfo{person}{Evropi Stefanidi}, \bibinfo{person}{Thomas Mildner},
  \bibinfo{person}{Yvonne Rogers}, {and} \bibinfo{person}{Jasmin Niess}.}
  \bibinfo{year}{2023}\natexlab{}.
\newblock \showarticletitle{SelVReflect: {A} guided {VR} experience fostering
  reflection on personal challenges}. In \bibinfo{booktitle}{\emph{Proceedings
  of the 2023 {ACM} {CHI} Conference on Human Factors in Computing Systems}}.
  \bibinfo{publisher}{{ACM}}, \bibinfo{address}{New York, NY},
  \bibinfo{pages}{323:1--323:17}.
\newblock
\urldef\tempurl%
\url{https://doi.org/10.1145/3544548.3580763}
\showDOI{\tempurl}


\bibitem[\protect\citeauthoryear{Weber, Mahmood, Ahmadi, von Jan, Ludwig, and
  Wieching}{Weber et~al\mbox{.}}{2023}]%
        {DBLP:journals/icom/WeberMAJLW23}
\bibfield{author}{\bibinfo{person}{Philip Weber}, \bibinfo{person}{Faisal
  Mahmood}, \bibinfo{person}{Michael Ahmadi}, \bibinfo{person}{Vanessa von
  Jan}, \bibinfo{person}{Thomas Ludwig}, {and} \bibinfo{person}{Rainer
  Wieching}.} \bibinfo{year}{2023}\natexlab{}.
\newblock \showarticletitle{Fridolin: Participatory Design and Evaluation of a
  Nutrition Chatbot for Older aAults}.
\newblock \bibinfo{journal}{\emph{i-com}} \bibinfo{volume}{22},
  \bibinfo{number}{1} (\bibinfo{year}{2023}), \bibinfo{pages}{33--52}.
\newblock
\urldef\tempurl%
\url{https://doi.org/10.1515/ICOM-2022-0042}
\showDOI{\tempurl}


\bibitem[\protect\citeauthoryear{Weimann, Schlieter, and Brendel}{Weimann
  et~al\mbox{.}}{2022}]%
        {weimann2022virtual}
\bibfield{author}{\bibinfo{person}{Thure~Georg Weimann},
  \bibinfo{person}{Hannes Schlieter}, {and} \bibinfo{person}{Alfred~Benedikt
  Brendel}.} \bibinfo{year}{2022}\natexlab{}.
\newblock \showarticletitle{Virtual coaches: Background, theories, and future
  research directions}.
\newblock \bibinfo{journal}{\emph{Business \& Information Systems Engineering}}
  \bibinfo{volume}{64}, \bibinfo{number}{4} (\bibinfo{year}{2022}),
  \bibinfo{pages}{515--528}.
\newblock


\bibitem[\protect\citeauthoryear{Whitworth, Kimsey-House, Kimsey-House, and
  Sandahl}{Whitworth et~al\mbox{.}}{1998}]%
        {Whitworth1998}
\bibfield{author}{\bibinfo{person}{Laura Whitworth}, \bibinfo{person}{Karen
  Kimsey-House}, \bibinfo{person}{Henry Kimsey-House}, {and}
  \bibinfo{person}{Phillip Sandahl}.} \bibinfo{year}{1998}\natexlab{}.
\newblock \bibinfo{booktitle}{\emph{Co-active coaching: New skills for coaching
  people toward success in work and life}}.
\newblock \bibinfo{publisher}{Nicholas Brealey}, \bibinfo{address}{Boston, MA}.
\newblock


\bibitem[\protect\citeauthoryear{Witherspoon and White}{Witherspoon and
  White}{1996}]%
        {witherspoon1996executive}
\bibfield{author}{\bibinfo{person}{Robert Witherspoon} {and}
  \bibinfo{person}{Randall~P White}.} \bibinfo{year}{1996}\natexlab{}.
\newblock \showarticletitle{Executive coaching: A continuum of roles}.
\newblock \bibinfo{journal}{\emph{Consulting Psychology Journal: Practice and
  Research}} \bibinfo{volume}{48}, \bibinfo{number}{2} (\bibinfo{year}{1996}),
  \bibinfo{pages}{124--133}.
\newblock
\urldef\tempurl%
\url{https://doi.org/10.1037/1061-4087.48.2.124}
\showDOI{\tempurl}


\bibitem[\protect\citeauthoryear{Wlasak, Zwanenburg, and Paton}{Wlasak
  et~al\mbox{.}}{2023}]%
        {Wlasak2023Supporting}
\bibfield{author}{\bibinfo{person}{Wendy Wlasak}, \bibinfo{person}{Sander~Paul
  Zwanenburg}, {and} \bibinfo{person}{Chris Paton}.}
  \bibinfo{year}{2023}\natexlab{}.
\newblock \showarticletitle{Supporting autonomous motivation for physical
  activity with chatbots during the {COVID}-19 pandemic: Factorial experiment}.
\newblock \bibinfo{journal}{\emph{{JMIR} Formative Research}}
  \bibinfo{volume}{7} (\bibinfo{year}{2023}), \bibinfo{pages}{e38500}.
\newblock
\urldef\tempurl%
\url{https://doi.org/10.2196/38500}
\showDOI{\tempurl}


\bibitem[\protect\citeauthoryear{Wolfbauer, Pammer{-}Schindler, Maitz, and
  Ros{\'{e}}}{Wolfbauer et~al\mbox{.}}{2022}]%
        {DBLP:journals/tlt/WolfbauerPMR22}
\bibfield{author}{\bibinfo{person}{Irmtraud Wolfbauer},
  \bibinfo{person}{Viktoria Pammer{-}Schindler}, \bibinfo{person}{Katharina
  Maitz}, {and} \bibinfo{person}{Carolyn~P. Ros{\'{e}}}.}
  \bibinfo{year}{2022}\natexlab{}.
\newblock \showarticletitle{A Script for Conversational Reflection Guidance:
  {A} Field Study on Developing Reflection Competence With Apprentices}.
\newblock \bibinfo{journal}{\emph{{IEEE} Transactions on Learning
  Technologies}} \bibinfo{volume}{15}, \bibinfo{number}{5}
  (\bibinfo{year}{2022}), \bibinfo{pages}{554--566}.
\newblock
\urldef\tempurl%
\url{https://doi.org/10.1109/TLT.2022.3207226}
\showDOI{\tempurl}


\bibitem[\protect\citeauthoryear{Wood, Linley, Maltby, Baliousis, and
  Joseph}{Wood et~al\mbox{.}}{2008}]%
        {Wood2008Authentic}
\bibfield{author}{\bibinfo{person}{Alex~M. Wood}, \bibinfo{person}{P.~Alex
  Linley}, \bibinfo{person}{John Maltby}, \bibinfo{person}{Michael Baliousis},
  {and} \bibinfo{person}{Stephen Joseph}.} \bibinfo{year}{2008}\natexlab{}.
\newblock \showarticletitle{The authentic personality: A theoretical and
  empirical conceptualization and the development of the authenticity scale}.
\newblock \bibinfo{journal}{\emph{Journal of Counseling Psychology}}
  \bibinfo{volume}{55}, \bibinfo{number}{3} (\bibinfo{year}{2008}),
  \bibinfo{pages}{385--399}.
\newblock
\urldef\tempurl%
\url{https://doi.org/10.1037/0022-0167.55.3.385}
\showDOI{\tempurl}


\bibitem[\protect\citeauthoryear{Xygkou, Siriaraya, Covaci, Prigerson,
  Neimeyer, Ang, and She}{Xygkou et~al\mbox{.}}{2023}]%
        {Xygkou2023Conversation}
\bibfield{author}{\bibinfo{person}{Anna Xygkou}, \bibinfo{person}{Panote
  Siriaraya}, \bibinfo{person}{Alexandra Covaci}, \bibinfo{person}{Holly~Gwen
  Prigerson}, \bibinfo{person}{Robert Neimeyer}, \bibinfo{person}{Chee~Siang
  Ang}, {and} \bibinfo{person}{Wan-Jou She}.} \bibinfo{year}{2023}\natexlab{}.
\newblock \showarticletitle{The "conversation" about loss: Understanding how
  chatbot technology was used in supporting people in grief}. In
  \bibinfo{booktitle}{\emph{Proceedings of the 2023 {ACM} {CHI} Conference on
  Human Factors in Computing Systems}}. \bibinfo{publisher}{{ACM}},
  \bibinfo{address}{New York, NY}, \bibinfo{pages}{646:1--646:15}.
\newblock
\urldef\tempurl%
\url{https://doi.org/10.1145/3544548.3581154}
\showDOI{\tempurl}


\bibitem[\protect\citeauthoryear{Zhao, Liu, Qiu, and Luo}{Zhao
  et~al\mbox{.}}{2020}]%
        {DBLP:journals/computer/ZhaoLQL20}
\bibfield{author}{\bibinfo{person}{Wenbing Zhao}, \bibinfo{person}{Xiongyi
  Liu}, \bibinfo{person}{Tie Qiu}, {and} \bibinfo{person}{Xiong Luo}.}
  \bibinfo{year}{2020}\natexlab{}.
\newblock \showarticletitle{Virtual avatar-based life coaching for children
  with autism spectrum disorder}.
\newblock \bibinfo{journal}{\emph{Computer}} \bibinfo{volume}{53},
  \bibinfo{number}{2} (\bibinfo{year}{2020}), \bibinfo{pages}{26--34}.
\newblock
\urldef\tempurl%
\url{https://doi.org/10.1109/MC.2019.2915979}
\showDOI{\tempurl}


\end{thebibliography}

\end{document}